\documentclass[pre,aps,eqsecnum,preprint]{revtex4}
\usepackage{graphicx}
\usepackage{amsmath}
\usepackage{epsfig}
\usepackage{color}
\definecolor{red}{rgb}{1,0,0}
\definecolor{green}{rgb}{0.13,0.55,0.13}
\definecolor{blue}{rgb}{0,0,1}

\begin{document}

\title{Two-dimensional critical systems with mixed boundary conditions: Exact Ising results from conformal invariance and boundary-operator expansions} 

\author{ T. W. Burkhardt$^{1 }$ and E. Eisenriegler$^{2}$ }

\affiliation{
$^1$Department of Physics, Temple University, Philadelphia, PA 19122, USA\\
$^2$Theoretical Soft Matter and Biophysics, Institute of Complex Systems, Forschungszentrum J\"ulich, D-52425 J\"ulich, Germany
}
\date{\today}
\begin{abstract}
With conformal-invariance methods, Burkhardt, Guim, and Xue studied the critical Ising model, defined on the upper half plane $y>0$ with different boundary conditions $a$ and $b$ on the negative and positive $x$ axes. For $ab=-+$ and $f+$, they determined the one and two-point averages of the spin $\sigma$ and energy $\epsilon$. Here $+$, $-$, and $f$ stand for spin-up, spin-down, and free-spin boundaries, respectively. The case $+-+-+\dots$, where the boundary conditions switch between $+$ and $-$ at arbitrary points, $\zeta_1$, $\zeta_2$, $\dots$ on the $x$ axis was also analyzed. 

In this paper the alternating boundary conditions $+f+f+\dots$ and the case $-f+$ of three different boundary conditions are considered. Exact results for  the one and two-point averages of $\sigma$, $\epsilon$, and the stress tensor $T$ are derived. Using the results for $\langle T\rangle$, the critical Casimir interaction with the boundary of a wedge-shaped inclusion is analyzed for mixed boundary conditions. 

The paper also includes a comprehensive discussion of boundary-operator expansions in two-dimensional critical systems with mixed boundary conditions. Two types of expansions - away from switching points of the boundary condition and at switching points - are considered. The asymptotic behavior of two-point averages is expressed in terms of one-point averages with the help of the expansions. We also consider the strip geometry with mixed boundary conditions and derive the distant-wall corrections to one-point averages near one edge due to the other edge using the boundary-operator expansions. The predictions of the boundary-operator expansions are consistent with exact results for Ising systems.
\end{abstract}

\maketitle
 
\section{Introduction} \label{intro}

The conformal-invariance approach of Belavin et al. \cite{BPZ,CardyD-L} determines the universal bulk properties, including critical indices and correlation functions, of an infinite class of two-dimensional systems at the critical point. Cardy \cite{Cardyscp} extended the approach to semi-infinite two-dimensional critical systems with a uniform boundary condition, such as free or fixed boundary spins. Cardy \cite{Cardytab} and Burkhardt and Xue \cite{TWBX} made a further extension to semi-infinite critical systems with mixed, piecewise-uniform boundary conditions. 

Of systems with mixed boundary conditions, the Ising model has received the most attention. For the Ising model on the upper half half plane $y>0$, with boundary conditions $a$ and $b$ on the negative and positive $x$ axes, the one and two-point averages  $\langle \sigma\rangle$,  $\langle \sigma_1\sigma_2\rangle$, $\langle \epsilon\rangle$, $\langle \epsilon_1\epsilon_2\rangle$, and $\langle \sigma_1\epsilon_2\rangle$ were derived by Burkhardt, Guim, and Xue in Refs. \cite{TWBG1,TWBX,TWBG2} for $ab= -+$ and $f+$. Here $\sigma$ and $\epsilon$ are the spin and energy operators, and $+$, $-$, and $f$ stand for spin-up, spin-down, and free-spin boundary conditions, respectively.  The case of alternating boundary conditions $+-+-+\dots$, which switch between $+$ and $-$ at arbitrary points $\zeta_1$, $\zeta_2$, $\dots$ on the $x$ axis is considered in \cite{TWBG2}.

In the first half of this paper the Ising model with alternating boundary conditions $+f+f+\dots$ and with three different boundaries $-f+$ is analyzed with conformal-invariance methods. Exact results  for the  one and two-point averages of $\sigma$, $\epsilon$, and the complex stress tensor $T(z)$ are obtained. 

The average stress tensor is of interest in connection with Casimir or fluctuation-induced interaction of particles immersed in a two-dimensional critical fluid or of a single particle with the linear boundary of the fluid \cite{EETWB,SMED,TWBEE,BEK}. For a two-dimensional critical system defined on the upper half plane with a uniform boundary condition on the $x$ axis, $\langle T(z)\rangle=0$, where $z=x+iy$. In the case of boundary condition $a$ for $x<\zeta_1$ and $b$ for $x>\zeta_1$, 
\begin{equation}
\langle T(z)\rangle_{ab}={t_{ab}\over(z-\zeta_1)^2}\,,\label{Tab}
\end{equation}
where the amplitude $t_{ab}=t_{ba}$ depends on the bulk universality class \cite{Cardytab,TWBX}. For the Ising model $t_{+-}={1\over 2}$, and  $t_{f+}=t_{f-}={1\over 16}$. 

For $aba$ and $abc$ boundaries, with changes in the boundary condition at points $\zeta_1$ and $\zeta_2$ on the $x$ axis,
\begin{eqnarray}
&&\langle T(z)\rangle_{aba}=t_{ab}\left({1\over z-\zeta_1}-{1\over z-\zeta_2}\right)^2\,,\label{Taba}\\
&&\langle T(z)\rangle_{abc}={t_{ab}\over (z-\zeta_1)^2}+{t_{bc}\over (z-\zeta_2)^2}+{t_{ac}-t_{ab}-t_{bc}\over(z-\zeta_1)(z-\zeta_2)}\,.\label{Tabc}
\end{eqnarray}
Since $t_{aa}=0$, Eq.~(\ref{Tabc}) reproduces Eq.(\ref{Taba}) for $c=a$.
Expressions (\ref{Taba}) and (\ref{Tabc}) are dictated by the requirements that $\langle T(z)\rangle$ scale as (length)$^{-2}$, diverge as in Eq.~(\ref{Tab}) for $z\to\zeta_1$ and  $z\to\zeta_2$, and reduce to the results for $aa$ and $ac$ boundary conditions, respectively, in the limit $\zeta_2\to\zeta_1$. Equation (\ref{Taba}) also follows from Eq.~(\ref{Tab}) and the transformation property (\ref{Ttransform}) of the stress tensor under the conformal transformation
\begin{equation}
z'=\zeta_1-\left({z-\zeta\over\Lambda^2}-{1\over\zeta_2-\zeta_1}\right)^{-1}\,,\label{abtoaba}
\end{equation}
which maps the $ba$ geometry onto $aba$. Here $\Lambda$ is an arbitrary constant with the dimensions of length.
 
In cases where the boundary condition changes at more than two points, for example, for $ababa$, $\langle T(z)\rangle$ is no longer uniquely determined by such elementary considerations, but the explicit form follows from the conformal-invariance approach, as shown below.

The paper is organized as follows: In Sec.~\ref{IsingConformal} the semi-infinite critical Ising model is studied with conformal-invariance methods for alternating $+f+f+\dots$ boundaries and in the case $-f+$ of three different boundary conditions. The exact one and two-point averages of the spin $\sigma$, energy $\epsilon$  and stress tensor $T$ are derived for these boundary conditions in Subsecs.~\ref{pfpetc} and \ref{mfp}. In Subsec.~\ref{wedge} we analyze the critical Casimir force on an infinite, wedge-shaped inclusion in the upper half plane, oriented perpendicular to the $x$ axis. For an $f$ boundary along the $x$ axis and $+$ and $-$ boundary conditions on the left and right edges of the wedge, the Casimir force reverses direction at a critical value of the apex angle.

The expansion of operators, such as $\sigma$ and $\epsilon$, near boundaries in terms of boundary operators has  been studied extensively for {\em uniform} boundary conditions \cite{Diehl,CardyLewellen,EEStap,Cardydistantwall}. In Sec.~\ref{secMBOE} a comprehensive analysis of boundary-operator expansions in two-dimensional critical systems with {\em mixed} boundary conditions is presented. Two types of expansions - away from switching points of the boundary condition and at switching points - are considered. The asymptotic form of two-point averages is expressed in terms of one-point  averages using the boundary-operator expansions. In another application of the expansions, to strips with mixed boundary conditions, we derive the distant-wall corrections to one-point averages near one edge due to the other, distant edge. All of the predictions for Ising systems based on the boundary-operator expansions are confirmed by comparison with exact results.

Section~\ref{concludingremarks} contains concluding remarks. 

\section{Exact Ising results from conformal invariance}\label{IsingConformal}
\subsection{Conformal differential equations}\label{confdiffeq}
In the conformal classification \cite{BPZ,CardyD-L} the spin $\sigma$ and energy $\epsilon$ of the Ising model are both degenerate at level 2. The bulk $n$-point average $\langle\sigma_1\dots\sigma_\ell\,\epsilon_{\ell+1}\dots\epsilon_{n}\rangle$ satisfies the $n$ partial differential equations 
\begin{equation}
\left[-{3\over 2(1+2\Delta_i)}\,{\partial^2\over\partial z_i^2}+\sum_{\substack{j=1\\ j\neq i}}^n\left({1\over z_{ij}}\,{\partial\over\partial z_j}+{\Delta_j\over z_{ij}^2}\right)\right]G^{(n)}(z_1,z_2,\dots,z_n)=0\,,\label{conformdiffeq}
\end{equation}
with $\Delta_k={1\over 16}$ for the spin and $\Delta_k={1\over 2}$ for the energy. Here $z_j=x_j+iy_j$ is the position of point $j$ in the complex plane, and $z_{ij}=z_i-z_j$. 

Burkhardt and Guim \cite{TWBG2} have discussed the solutions of Eq.~(\ref{conformdiffeq}) in the cases $\Delta_1=\Delta_2=\dots =\Delta_n={1\over 16}\;{\rm and}\;{1\over 2}$, corresponding to $\langle\sigma_1\dots\sigma_n\rangle$ and  $\langle\epsilon_1\dots\epsilon_n\rangle$, respectively. In the former case, they showed that for even $n$ there are $2^{n/2-1}$ linearly independent solutions of differential equations (\ref{conformdiffeq}) given by

\begin{eqnarray}
&&G_\sigma^{(n,\alpha)}(z_1,\dots, z_n)=(z_{12}z_{34}\dots z_{n-1,n})^{-1/8}\nonumber\\
&&\qquad\times\Bigg\{{1\over 2}\, \sum_{\tau_1=\pm 1}\sum_{\tau_3=\pm 1}\dots\sum_{\tau_{n-1}=\pm 1}S_\alpha(\tau_1,\tau_3,\dots,\tau_{n-1})\prod_{\substack{i<j \\ i,j\;{\rm odd}}}\xi_{ij}^{\tau_i\tau_j}\Bigg\}^{1/2},\label{Gnsigma}\\
&&\xi_{ij}=\left({z_{i,j}\,z_{i+1,j+1}\over z_{i,j+1}\,z_{i+1,j}}\right)^{1/4},\label{xidef}
\end{eqnarray}
where $\alpha= 1,2,\dots,2^{n/2-1}$.
The quantities $S_\alpha(\tau_1,\tau_3,\dots,\tau_{n-1})$ are the even operators $1$, $\tau_k\tau_\ell$ with $k<\ell$, $\tau_k\tau_\ell\tau_m\tau_n$ with $k<\ell<m<n$, etc., where $k,\ell,\dots$ take the values $1,3,5\dots,n-1$. For $n=2$ and 4 ,
\begin{eqnarray}
&& G_\sigma^{(2,1)}(z_1,z_2)=z_{12}^{-1/8}\,,\label{G2sigma}\\
&& G_\sigma^{(4,1)}(z_1,\dots,z_4)=(z_{12}z_{34})^{-1/8}\left(\xi_{13}+\xi_{13}^{-1}\right)^{1/2}\,,\label{G41sigma}\\
&& G_\sigma^{(4,2)}(z_1,\dots,z_4)=(z_{12}z_{34})^{-1/8}\left(\xi_{13}-\xi_{13}^{-1}\right)^{1/2}\,,\label{G42sigmasigma}\end{eqnarray}
and for $n=6$,
\begin{eqnarray}
&&G^{(6,\alpha)}_\alpha (z_1,\dots,z_6)\nonumber\\
&&\qquad=(z_{12}z_{34} z_{56})^{-1/8}\left(\xi_{13}\xi_{15}\xi_{35}+C_{\alpha1}\,{\xi_{13}\over\xi_{15}\xi_{35}}+C_{\alpha2}\,{\xi_{15}\over\xi_{13}\xi_{35}}+C_{\alpha3}\,{\xi_{35}\over\xi_{13}\xi_{15}}\right)^{1/2}\,,\label{G6a}
\end{eqnarray}
with matrix $C$ of coefficients 
\begin{equation}
C=\left(\begin{matrix} 1&1&1\\ 1&-1&-1\\ -1&1&-1\\ -1&-1&1\end{matrix}\right)\,.\label{G6b}
\end{equation}

For $\Delta={1\over 2}$, corresponding to the energy,  there appears to be only one physical solution of differential equation (\ref{conformdiffeq}), given by
\begin{eqnarray}
&& G_\epsilon^{(2)}(z_1,z_2)=z_{12}^{-1}\,,\label{G2eps}\\
&& G_\epsilon^{(4)}(z_1,\dots,z_4)=(z_{12}z_{34})^{-1}-(z_{13}z_{24})^{-1}+(z_{14}z_{23})^{-1},\label{G4eps}\\
&& G_\epsilon^{(n)}(z_1,\dots,z_n)={\rm Pf}^{(n)}{1\over z_{ij}}\,,\label{Gneps}
\end{eqnarray}
for $n=2$, 4, and general even $n$.
Here ${\rm Pf}^{(n)}A_{ij}$ denotes the Paffian of the $n\times n$ antisymmetric matrix with elements $A_{ij}$.

From these solutions Burkhardt and Guim \cite{TWBG2} constructed all the correlation functions $\langle\sigma_1\sigma_2\dots\sigma_n\rangle$ and $\langle\epsilon_1\epsilon_2\dots\epsilon_n\rangle$ both in the bulk and in the half space with uniform fixed and free-spin boundary conditions. In addition they derived the one and two-point averages of the spin and energy density in the half space with alternating $+-+-+\dots$ boundary conditions. 

\subsection{Boundary condition $+f+f+\dots$}
\label{pfpetc}
\subsubsection{General approach for alternating boundary conditions}
We begin by considering the correlations of an arbitrary primary operator $\phi(z,\bar z)$ in a semi-infinite critical system defined on the upper half plane, with $ababa\dots$ boundary conditions, which switch between $a$ and $b$ at an even number $m$ of points $\zeta_1<\zeta_2<\dots<\zeta_m$ on the $x$ axis. For $-\infty<x<\zeta_1$ the boundary condition is $a$, for $\zeta_1<x<\zeta_2$ it is $b$, for $\zeta_2<x<\zeta_3$ it is $a$, etc. Results for odd number $m-1$  of $\zeta$'s are obtained by taking the limit $\zeta_m\to\infty$.

Following  \cite{Cardyscp,TWBX,TWBG1,TWBG2}, we express  the $n$-point correlation function of $\phi$ as
\begin{equation}
\langle\phi_1\dots\phi_n\rangle_{ababa\,\dots}={N(\zeta_1,\dots,\zeta_m,z_1,\bar  z_1,\dots,z_n,\bar z_n)\over D(\zeta_1,\dots,\zeta_m)}\,,\label{NoverD1}
\end{equation}
where the numerator $N$ satisfies the same differential equations in the $m+2n$ variables $(\zeta_1,\dots,\zeta_m,z_1,\bar  z_1,\dots,z_n,\bar z_n)$ as the bulk correlation function $\langle\psi_1\dots\psi_m\,\phi_{m+1}\dots\phi_{m+2n}\rangle_{\rm bulk}$ in the variables $(z_1,z_2,\dots z_{m+2n})$. In these differential equations the scaling index $\Delta_i$ for the operators $\phi_{m+1},\dots,\phi_{m+2n}$ is the usual bulk index $\Delta_\phi$. For the operators $\psi_1,\dots,\psi_m$, $\Delta_i=t_{ab}$, where $t_{ab}$ is the boundary index introduced in Eq.~(\ref{Tab}).
The denominator $D$ in Eq.~(\ref{NoverD1}) satisfies the same differential equations in the variables $\zeta_1,\dots,\zeta_m$ as  the bulk correlation function $\langle\psi_1\dots\psi_m\rangle_{\rm bulk}$ in the variables $z_1,\dots,z_m$, with $\Delta_i=t_{ab}$.
    
In the limit that all of the $n$ points are translated infinitely far to the left of $\zeta_1$ without changing their relative positions, $\langle\phi_1\dots\phi_n\rangle_{ababa\,\dots}$ reduces to the corresponding correlation function for a uniform boundary condition $a$. All of the correlation functions we consider are known for a uniform boundary condition. Thus, once the numerator $N$ in Eq.~(\ref{NoverD1}) has been determined, $D$ can be obtained from
\begin{equation}
D(\zeta_1,\dots,\zeta_m)\langle\phi_1\dots\phi_n\rangle_{a}=\lim_{X\to -\infty}N(\zeta_1,\dots,\zeta_m,z_1+X,\bar  z_1+X,\dots,z_n+X,\bar z_n+X)\,.\label{Dfromlimit}
\end{equation}
This procedure for determining $D$ is the simplest in practice, and it ensures that the correlation function (\ref{NoverD1}) for mixed boundary conditions is  correctly normalized.

For the Ising $n$-spin correlation function $\langle\sigma_1\dots\sigma_n\rangle_{+f+f+f+\,\dots}$, there is a simplifying feature. Both the bulk index $\Delta_\sigma$ and the boundary index $t_{+f}$ have the value ${1\over16}$, as mentioned just below Eqs.~(\ref{Tab}) and (\ref{conformdiffeq}). Thus, the numerator $N$ in Eq.~(\ref{NoverD1}) satisfies the same differential equations in the $m+2n$ variables $\zeta_1,\dots,\zeta_m,z_1,\bar  z_1,\dots,z_n,\bar z_n$ as the bulk $n$-spin correlation function in the variables $z_1,z_2,\dots z_{m+2n}$. This implies that $N$ is an appropriate linear combination of the $2^{m/2+n-1}$ functions $G_\sigma^{(m+2n,\alpha)}(\zeta_1,\dots,\zeta_m,z_1,\bar  z_1,\dots,z_n,\bar z_n)$ defined in Eq.~(\ref{Gnsigma}). Similarly $D$ is an appropriate linear combination of the $2^{m/2-1}$ functions $G_\sigma^{(m,\alpha)}(\zeta_1,\dots,\zeta_m)$. The linear combinations are determined by the requirement that $N/D$ reproduce the expected asymptotic behavior of  $\langle\sigma_1\dots\sigma_n\rangle_{+f+f+f+\,\dots}$ as any two of the $n$ points approach each other or as any of the points approaches the boundary line $y=0$ or approaches infinity parallel to the $x$ axis. The operator product expansion of closely spaced spin operators and the one point averages of $\sigma$ and $\epsilon$ in the presence of a homogeneous boundary are discussed in the next subsection. The general form of the operator expansion near a boundary point is considered in Sec.~\ref{secMBOE}.

\subsubsection{Operator product expansion}

To obtain correlation functions involving the energy $\epsilon$ from $\langle\sigma_1\dots\sigma_n\rangle_{+f+f+\,\dots}$, we make use of the operator-product expansion (OPE) of two closely spaced $\sigma$ operators. This and two other useful OPE's (see Eq.~(D6) of Ref.~\cite{EETWB}, Eqs.~(2.39), (2.47), (3.46), and (A1) of Ref.~\cite{EE}, and Eq.~(D.25) of Ref.~\cite{SMED}) are given by
\begin{eqnarray} 
&&\sigma(z_1,\bar z_1)\sigma(z_2,\bar z_2)=\vert z_{12}\vert^{-1/4}\nonumber\\
&&\qquad\times\left\{1-\textstyle{1\over 2}\,\vert z_{12}\vert\epsilon(z,\bar z) +{1\over 4}\,\left[z_{12}^2\,T(z)+\bar z_{12}^2\,\bar T(\bar z)\right] +\mathcal{O}\left(\vert z_{12}\vert^{3}\right)\right\}\,,\label{OPEsigsig}\\
&&\sigma(z_1,\bar z_1)\epsilon(z_2,\bar z_2)=-{1\over 2}\vert z_{12}\vert^{-1}\sigma(z,\bar z)\big[1+\mathcal{O}\left(\vert z_{12}\vert\right)\big]\,,\label{OPEsigeps}\\
&&\epsilon(z_1,\bar z_1)\epsilon(z_2,\bar z_2)=\vert z_{12}\vert^{-2}\left\{1+2\left[z_{12}^2\,T(z)+\bar z_{12}^2\,\bar T(\bar z)\right]+\mathcal{O}\left(\vert z_{12}\vert^{4}\right)\right\}\,,\label{OPEepseps}
\end{eqnarray}
where $z_{12}=z_1-z_2$ and $z={1\over 2}(z_1+z_2)$. 

In Eqs.~(\ref{OPEsigsig})-(\ref{OPEepseps}) we follow the convention of normalizing  $\sigma$ and $\epsilon$ so that the bulk pair correlation functions are $\langle\sigma_1\sigma_2\rangle_{\rm bulk}=\vert z_1-z_2\vert^{-1/4}$ and $\langle\epsilon_1\epsilon_2\rangle_{\rm bulk}=\vert z_1-z_2\vert^{-1}$. With this normalization the correlation functions in the upper half plane with a uniform boundary condition on the $x$ axis are given by \cite{Cardyscp,TWBX}
\begin{eqnarray}
&&\langle\sigma_1\sigma_2\rangle_{\rm fixed\,or\,free}=(4y_1y_2)^{-1/8}\left[{1\over\sqrt{\rho}}\pm\sqrt{\rho}\,\right]^{1/2}\,,\label{sigsiguniformbc1}\\[2mm]
&&\rho=\left[{(x_1-x_2)^2+(y_1-y_2)^2\over (x_1-x_2)^2+(y_1+y_2)^2}\right]^{1/2}\,,\label{sigsiguniformbc2}\\[2mm]
&&\langle\epsilon_1\epsilon_2\rangle_{\rm fixed\,or\,free}={1\over 4y_1y_2}+{1\over (x_1-x_2)^2+(y_1-y_2)^2}-{1\over (x_1-x_2)^2+(y_1+y_2)^2}\,.\label{epsepsuniformbc}
\end{eqnarray}
The upper and lower sign in Eq.~(\ref{sigsiguniformbc1}) holds for fixed and free boundary conditions, respectively, and Eq.~(\ref{epsepsuniformbc}) holds for both boundary conditions. 

Equations (\ref{sigsiguniformbc1})-(\ref{epsepsuniformbc}), the property $\langle\sigma_1\sigma_2\rangle\to \langle\sigma_1\rangle\langle\sigma_2\rangle$ for $x_2\to\infty$, and its analog for $\langle\epsilon_1\epsilon_2\rangle$ imply the one-point
averages
\begin{eqnarray}
&&\langle\sigma\rangle_{\rm fixed}=\pm \left({2\over y}\right)^{1/8}\,,\label{sigfixed}\\
&&\langle\epsilon\rangle_{\rm fixed\,or\,free}=\mp {1\over 2y}\,.\label{epsfixedfree}
\end{eqnarray}
In Eq.~(\ref{sigfixed}) the upper and lower signs  correspond to spin-up and spin-down boundary conditions, respectively, and in Eq.~(\ref{epsfixedfree}) they correspond to fixed and free boundaries. Equation (\ref{epsfixedfree}), including the $\mp$ sign, also follows directly from $\langle\sigma_1\sigma_2\rangle$ in Eq.~(\ref{sigsiguniformbc1}) and the short-distance expansion of $\sigma_1\sigma_2$ in Eq.~(\ref{OPEsigsig}).

\subsubsection{Average spin  $\langle\sigma\rangle_{+f+f+\,\dots}$}
Here we consider the average spin at point $(x,y)$ of the critical Ising model defined on the upper half plane, with alternating boundary conditions $+f+f+\dots$, which switch at an even number $m$ of points $\zeta_1<\zeta_2<\dots<\zeta_m$ on the $x$ axis. 
According to the discussion below Eq.~(\ref{NoverD1}), $\langle\sigma\rangle_{+f+f+\,\dots}=N/D$, where $N$ and $D$ are appropriate linear combinations of the functions $G_\sigma^{(m+2,\alpha)}(\zeta_1,\dots,\zeta_m,z,\bar  z,)$ and  $G_\sigma^{(m,\alpha)}(\zeta_1,\dots,\zeta_m)$, respectively. 
The linear combinations turn out to be particularly simple, consisting only of the function with $\alpha=1$.
In this subsection we argue that
\begin{eqnarray}
&&\langle\sigma\rangle_{+f+f+\,\dots}=\left({i\over 4}\right)^{1/8}\,{G_\sigma^{(2+m,1)}(\zeta_1,\dots,\zeta_m,z,\bar  z)\over G_\sigma^{(m,1)}(\zeta_1,\dots,\zeta_m)}\,.
\label{NoverD2}
\end{eqnarray}
has the correct asymptotic behavior for  $y\to 0$, whereas other linear combinations do not. We now show this explicitly for $m=4$, with an argument which is easily extended to other even $m$.

Making the replacement $(z_1,\dots,z_6)\to(\zeta_1,\dots,\zeta_4,z,\bar z)$ and combining Eqs.~ (\ref{xidef}), (\ref{G41sigma}), (\ref{G6a}), 
(\ref{G6b}), and (\ref{NoverD2}), we obtain 
\begin{eqnarray}
&&\langle\sigma\rangle_{+f+f+}=\nonumber\\&&\quad=\left({1\over 8y}\right)^{1/8}\left[{\xi_{13}\xi_{15}\xi_{35}+\xi_{13}(\xi_{15}\xi_{35})^{-1}+\xi_{15}(\xi_{13}\xi_{35})^{-1}+{\xi_{35}(\xi_{13}\xi_{15})^{-1}}\over \xi_{13}+\xi_{13}^{-1}}\right]^{1/2},\label{NoverD3a}\\
&&\xi_{13}=\left[{(\zeta_1-\zeta_3)(\zeta_2-\zeta_4)\over(\zeta_1-\zeta_4)(\zeta_2-\zeta_3)}\right]^{1/4},\label{NoverD3b}\\
&& \xi_{15}=\left[{(\zeta_1-z)(\zeta_2-\bar z)\over(\zeta_1-\bar z)(\zeta_2-z)}\right]^{1/4}=e^{i(\varphi_1-\varphi_2))/2},
\label{NoverD3c}\\
&& \xi_{35}=\left[{(\zeta_3-z)(\zeta_4-\bar z)\over(\zeta_3-\bar z)(\zeta_4-z)}\right]^{1/4}=e^{i(\varphi_3-\varphi_4)/2},\label{NoverD3d}
\end{eqnarray}
consistent with Eqs.~(\ref{G41sigma}), (\ref{G6a}) and (\ref{G6b}) for $\alpha=1$.                        
Here $z-\zeta_j=|z-\zeta_j|e^{i\varphi_j}$, and $\varphi_j$ is the angle which a line from $\zeta_j$ to $z=x+iy$ in the complex plane forms with the $x$ axis. 

To check that Eqs.~(\ref{NoverD3a})-(\ref{NoverD3d}) satisfy the $+f+f+$ boundary condition, first suppose that  $x<\zeta_1$. Then, in the limit $y\to 0$ all four angles $\varphi_1,\dots,\varphi_4$ approach $\pi$, implying $\xi_{15}\to 1$ and $\xi_{35}\to 1$. Thus, the square bracket in Eq.~(\ref{NoverD3a}) approaches 2, consistent with the spin-up  boundary condition (\ref{sigfixed}) for $x<\zeta_1$.

Now suppose that  $\zeta_1<x<\zeta_2$. Then, in the limit $y\to 0$, $\varphi_1$ approaches $0$, and $\varphi_2$, $\varphi_3$ and $\varphi_4$, all approach $\pi$, implying $\xi_{15}\to -i$, $\xi_{35}\to 1$. Thus, the square bracket in Eq.~(\ref{NoverD2}) vanishes, consistent with the free spin boundary condition for $\zeta_1<x<\zeta_2$. 

Considering the two remaining possibilities $\zeta_2<x<\zeta_3$ and $x>\zeta_4$ in the same way, we confirm the full consistency of Eqs.~(\ref{NoverD3a})-(\ref{NoverD3d}) with the  $+f+f+$ boundary condition.

\subsubsection{Correlation function $\langle\sigma_1\dots\sigma_n\rangle_{+f+f+\,\dots}$}
Now we turn to the $n$-spin correlation function of the semi-infinite critical Ising with the same alternating boundary condition $+f+f+\dots$ as in the preceding subsection. According to the discussion below Eq.~(\ref{NoverD1}), $\langle\sigma_1\dots\sigma_n\rangle_{+f+f+\,\dots}=N/D$, where $N$ and $D$ are appropriate linear combinations of the functions 
$G_\sigma^{(m+2n,\alpha)}(\zeta_1,\dots,\zeta_m,z_1,\bar  z_1,\dots,z_n,\bar z_n)$ and  $G_\sigma^{(m,\alpha)}(\zeta_1,\dots,\zeta_m)$, respectively.
As in the preceding subsection, we find that the linear combinations only involve the $G_\sigma$ with $\alpha=1$. Choosing the multiplicative constant for consistency with the normalization (\ref{sigfixed}) leads to
\begin{eqnarray}
&&\langle\sigma_1\dots\sigma_n\rangle_{+f+f+\,\dots}=\left({i\over 4}\right)^{n/8}\,{G_\sigma^{(m+2n,1)}(\zeta_1,\dots,\zeta_m,z_1,\bar  z_1,\dots,z_n,\bar z_n)\over G_\sigma^{(m,1)}(\zeta_1,\dots,\zeta_m)}\,,\label{NoverD4}
\end{eqnarray}

Beginning with Eq.~(\ref{NoverD4}), we have derived the one and two-point averages $\langle\sigma\rangle$, $\langle\epsilon\rangle$, $\langle\sigma_1\sigma_2\rangle$, $\langle\sigma_1\epsilon_2\rangle$, and $\langle\epsilon_1\epsilon_2\rangle$ both for $+f+$ and $+f+f+$ boundary conditions. The results are given in the next two paragraphs. The correlation functions  $\langle\epsilon\rangle$, $\langle\sigma_1\epsilon_2\rangle$, and $\langle\epsilon_1\epsilon_2\rangle$ were obtained from the 2, 3, and 4 spin correlation functions given by Eq.~(\ref{NoverD4}) on letting pairs of spins approach each other and  comparing with the operator product expansion (\ref{OPEsigsig}).\\

\paragraph{Results for $+f+$ boundary conditions}
For the boundary condition of up spins for $x_1<\zeta$, free spins for $\zeta_1<x<\zeta_2$, and up spins for $x>\zeta_2$,
\begin{eqnarray}
&&\langle\sigma_1\rangle_{+f+}=\left({2\over y_1}\right)^{1/8}\sqrt{\cos\left(\textstyle{1\over 2}\gamma_{1,1}\right)}\,.\label{sigpfp}\\[3mm]
&& \langle\epsilon_1\rangle_{+f+}=-{1\over 2y_1}\,\cos\gamma_{1,1}\,.\label{epspfp}
\end{eqnarray}
\begin{eqnarray}
&&\langle\sigma_1\sigma_2\rangle_{+f+}=\left({1\over 4y_1y_2}\right)^{1/8}\left[{1\over\sqrt{\rho}}\cos\left(\textstyle{1\over 2}\gamma_{1,1}-\textstyle{1\over 2}\gamma_{2,1}\right)+\sqrt{\rho}\cos\left(\textstyle{1\over 2}\gamma_{1,1}+\textstyle{1\over 2}\gamma_{2,1}\right)\right]^{1/2}.\label{sigsigpfp}\\[3mm]
&&\langle\sigma_1\epsilon_2\rangle_{+f+}=-
{1\over 2}\left({2\over y_1}\right)^{1/8}\left({1\over 2y_2}\right)\bigg[{1\over\rho}\,\cos\left(\textstyle{1\over 2}\gamma_{1,1}-\gamma_{2,1}\right)
\nonumber\\  &&\qquad\qquad\qquad\qquad\qquad+\rho\cos\left(\textstyle{1\over 2}\gamma_{1,1}+\gamma_{2,1}\right)\bigg] \Big/ \sqrt{\cos\left(\textstyle{1\over 2}\gamma_{1,1}\right)}\,.\label{sigepspfp}\\[3mm]
	&&\langle\epsilon_1\epsilon_2\rangle_{+f+}=-{1\over 8y_1y_2}\bigg[\left(1-{2\over\rho^2}\right)\cos(\gamma_{1,1}-\gamma_{2,1})\nonumber\\
&&\qquad\qquad\qquad\qquad\qquad +\left(1-2\rho^2\right)\cos(\gamma_{1,1}+\gamma_{2,1})\bigg]\,.\label{epsepspfp}
\end{eqnarray}
Here 
\begin{eqnarray}
&&\rho=\left[{(x_1-x_2)^2+(y_1-y_2)^2\over(x_1-x_2)^2+(y_1+y_2)^2}\right]^{1/2},\label{rhodef2}\\[2mm]
&& e^{i\gamma_{k,\ell}}={z_k-\zeta_{\ell+1}\over z_k-\zeta_\ell}\left\vert{z_k-\zeta_\ell\over z_k-\zeta_{\ell+1}}\right\vert
 ={(x_k-\zeta_\ell)(x_k-\zeta_{\ell+1})+y_k^2+iy_k(\zeta_{\ell+1}-\zeta_\ell)\over\sqrt{\left[(x_k-\zeta_\ell)(x_k-\zeta_{\ell+1})+y_k^2\right]^2+y_k^2(\zeta_{\ell+1}-\zeta_\ell)^2}}\,,
 \label{gammakelldef1}\nonumber\\   &&\\
&&\gamma_{k,\ell}={\rm arg}\left({x_k-\zeta_{\ell+1}+iy_k\over x_k-\zeta_\ell+iy_k}\right)\,.                                                                                                                                                                                                                                                                                              \label{gammakelldef2}
\end{eqnarray}

As a check on predictions (\ref{sigpfp})-(\ref{epsepspfp}), we note that in the limit $\zeta_1\to -\infty$, $\zeta_2\to 0$, they correctly reproduce the findings of Burkhardt and Xue \cite{TWBX} for free and fixed spins on the negative and positive $x$ axes, respectively. (Caution: An $ab$ boundary in our notation, corresponds to a  $ba$ boundary in the notation of Ref.~\cite{TWBX}.) Conversely, Eqs.~(\ref{sigpfp})-(\ref{epsepspfp}) may be derived from the results of Ref. \cite{TWBX} using the transformation properties of correlation functions under the conformal mapping (\ref{abtoaba}) of the $f+$ geometry onto the $+f+$ geometry.

It is straightforward to express predictions (\ref{sigpfp})-(\ref{epsepspfp}) entirely 
in terms of Cartesion coordinates. Since $y_k>0$ and $\zeta_{\ell+1}>\zeta_\ell$, the quantity $e^{i\gamma_{k,\ell}}$ in Eq.~(\ref{gammakelldef1}) has a positive imaginary part. Thus, $0<\gamma_{k,\ell}<\pi$, so that
\begin{eqnarray}
&&\cos\left(\textstyle{1\over 2}\gamma_{k,\ell}\right)=\textstyle{1\over\sqrt 2}\left(1+\cos\gamma_{k,\ell}\right)^{1/2}\,,\label{cosgammakellover2}\\
&&\sin\left(\textstyle{1\over 2}\gamma_{k,\ell}\right)=\textstyle{1\over\sqrt 2}\left(1-\cos\gamma_{k,\ell}\right)^{1/2}\,.\label{singammakellover2}
\end{eqnarray}
Substituting these relations, along with 
\begin{eqnarray}
&&\cos\gamma_{k,\ell}={(x_k-\zeta_\ell)(x_k-\zeta_{\ell+1})+y_k^2\over\sqrt{\left[(x_k-\zeta_\ell)(x_k-\zeta_{\ell+1})+y_k^2\right]^2+y_k^2(\zeta_{\ell+1}-\zeta_\ell)^2}}\,,\label{cosgammakell}\\
&&\sin\gamma_{k,\ell}={y_k(\zeta_{\ell+1}-\zeta_\ell)\over\sqrt{\left[(x_k-\zeta_\ell)(x_k-\zeta_{\ell+1})+y_k^2\right]^2+y_k^2(\zeta_{\ell+1}-\zeta_\ell)^2}}\,,\label{singammakell}
\end{eqnarray}
and the definition (\ref{rhodef2}) of $\rho$ 
in Eqs.~(\ref{sigpfp})-(\ref{epsepspfp}) leads to expressions in terms of Cartesian coordinates.\\

\paragraph{Results for $+f+f+$ boundary conditions}
For the $+f+f+$ boundary with changes at $\zeta_1,\dots,\zeta_4$, Eq.~(\ref{NoverD4}) and the same general procedure as in the preceding subsection lead to
\begin{eqnarray}
&&\langle\sigma_1\rangle_{+f+f+}=\left({2\over y_1}\right)^{1/8}\sqrt{{\cos\left(\textstyle{1\over 2}\gamma_{1,1}-\textstyle{1\over 2}\gamma_{1,3}\right)+\chi^2\cos\left(\textstyle{1\over 2}\gamma_{1,1}+\textstyle{1\over 2}\gamma_{1,3}\right)\over 1+\chi^2}}\,.\label{sigpfpfp}\nonumber\\\\[3mm]
&& \langle\epsilon_1\rangle_{+f+f+}=-{1\over 2y_1}\,{\cos\left(\gamma_{1,1}-\gamma_{1,3}\right)+\chi^2\cos\left(\gamma_{1,1}+\gamma_{1,3}\right)\over 1+\chi^2}\,.\label{epspfpfp}
\end{eqnarray}
\begin{eqnarray}
&&\langle\sigma_1\sigma_2\rangle_{+f+f+}=\left({1\over 4y_1y_2}\right)^{1/8}\bigg[\rho\cos\Big(\textstyle{1\over 2}(\gamma_{1,1}+\gamma_{2,1}-\gamma_{1,3}-\gamma_{2,3})\Big)\nonumber\\
&&\qquad\qquad +\chi^2 \cos\Big(\textstyle{1\over 2}(\gamma_{1,1}-\gamma_{2,1}+\gamma_{1,3}-\gamma_{2,3})\Big)+\cos\bigg(\textstyle{1\over 2}(\gamma_{1,1}-\gamma_{2,1}-\gamma_{1,3}+\gamma_{2,3})\bigg)\nonumber\\
&&\qquad\qquad +\rho\,\chi^2 \cos\Big(\textstyle{1\over 2}(\gamma_{1,1}+\gamma_{2,1}+\gamma_{1,3}+\gamma_{2,3})\bigg)\bigg]^{1/2}\bigg/\Big[4\sqrt{\rho}\, \left(1+\chi^2\right)\Big]^{1/2}.\label{sigsigpfpfp}\\[3mm]
&&\langle\sigma_1\epsilon_2\rangle_{+f+f+}=-\left({2\over y_1}\right)^{1/8}{1\over 2y_2}\nonumber\\
&&\qquad\qquad\times \bigg[\rho^2\cos\Big(\textstyle{1\over 2}\gamma_{1,1}-\textstyle{1\over 2}\gamma_{1,3}+\gamma_{2,1}-\gamma_{2,3}\Big)+\chi^2\cos\Big(\textstyle{1\over 2}\gamma_{1,1}+\textstyle{1\over 2}\gamma_{1,3}-\gamma_{2,1}-\gamma_{2,3}\Big)\nonumber\\
&&\qquad\qquad+\cos\Big(\textstyle{1\over 2}\gamma_{1,1}-\textstyle{1\over2}\gamma_{1,3}-\gamma_{2,1}+\gamma_{2,3}\Big)+\rho^2\chi^2\cos\Big(\textstyle{1\over 2}\gamma_{1,1}+\textstyle{1\over 2}\gamma_{1,3}+\gamma_{2,1}+\gamma_{2,3}\Big)\bigg]\nonumber\\ 
&&\qquad\qquad\times\bigg\{2\rho^2\left(1+\chi^2\right)\Big[\cos\left(\textstyle{1\over 2}\gamma_{1,1}-\textstyle{1\over 2}\gamma_{1,3}\right)+\chi^2\cos\left(\textstyle{1\over 2}\gamma_{1,1}+\textstyle{1\over 2}\gamma_{1,3}\right)\Big]\bigg\}^{-1/2}.
\label{sigepspfpfp}
\end{eqnarray}
\begin{eqnarray}
&&\langle\epsilon_1\epsilon_2\rangle_{+f+f+}={1\over 8y_1y_2 \rho^2(1+\chi^2)^2}\nonumber\\
&&\qquad\qquad\times \bigg\{\rho^2\left(-1+2\rho^2+2\rho^2\chi^2\right)\cos\left(\gamma_{1,1}- \gamma_{1,3}+\gamma_{2,1}-\gamma_{2,3}\right)\nonumber\\ 
&&\qquad\qquad +\chi^2\left(2+2\chi^2-\rho^2\chi^2\right)\cos\left(\gamma_{1,1}+\gamma_{1,3}-\gamma_{2,1}-\gamma_{2,3}\right)\nonumber\\ [2mm]
&&\qquad\qquad +\left(-\rho^2+2+2\chi^2\right)\cos\left(\gamma_{1,1}-\gamma_{1,3}-\gamma_{2,1}+\gamma_{2,3}\right)\nonumber\\[2mm] 
&&\qquad\qquad +\rho^2\chi^2\left(2\rho^2-\chi^2+2\rho^2\chi^2\right)\cos\left(\gamma_{1,1}+\gamma_{1,3}+\gamma_{2,1}+\gamma_{2,3}\right)\nonumber\\[2mm]
&&\qquad\qquad -\rho^2\chi^2\Big[\cos\left(-\gamma_{1,1}+\gamma_{1,3}+\gamma_{2,1}+\gamma_{2,3}\right)+\cos\left(\gamma_{1,1}-\gamma_{1,3}
+\gamma_{2,1}+\gamma_{2,3}\right)\nonumber\\
&&\qquad\qquad +\cos\left(\gamma_{1,1}+\gamma_{1,3}-\gamma_{2,1}+\gamma_{2,3}\right)+\cos\left(\gamma_{1,1}+\gamma_{1,3} +\gamma_{2,1}-\gamma_{2,3}\right)\Big]\bigg\}.\label{epsepspfpfp}
\end{eqnarray}
Here $\rho$ and $\gamma_{k,\ell}$ are the same as in Eqs.~(\ref{rhodef2})-(\ref{gammakelldef2}), and
\begin{equation}
\chi=\left[{(\zeta_1-\zeta_3)(\zeta_2-\zeta_4)\over(\zeta_1-\zeta_4)(\zeta_2-\zeta_3)}\right]^{1/4}.\,\label{chidef}
\end{equation}

It is simple to check the consistency of Eq.~(\ref{sigpfpfp}) for $\langle\sigma_1\rangle_{+f+f+}$ and our earlier result (\ref{NoverD3a})-(\ref{NoverD3d}),  with $\gamma_{1,1}=\varphi_2-\varphi_1$, $\gamma_{1,2}=\varphi_4-\varphi_3$, and $\chi=\xi_{13}$.

 \subsubsection{Average stress tensors $\langle T(z)\rangle_{+f+f+\,\dots}$ and $\langle T(z)\rangle_{+-+-+\,\dots}$} 
In the presence of mixed boundary conditions the average stress tensor does not vanish and appears explicitly in the conformal Ward identity and the differential equations for correlation functions \cite{Cardytab,TWBX}. Thus, while the numerator in expression (\ref{NoverD2}) for $\langle\sigma\rangle_{+f+f+\,\dots}$ obeys the differential equations with bulk-like form 
\begin{eqnarray}
&&\left[-{4\over 3}\,{\partial^2\over\partial z^2}+{1\over z-\bar z}\,{\partial\over\partial\bar z}+{1/16\over(z-\bar z)^2} +\sum_{j=1}^m\left({1\over z-\zeta_j}\,{\partial\over\partial \zeta_j}+{1/16\over(z-\zeta_j)^2}\right)\right]G_\sigma^{(2+m,1)}(\zeta_1,\dots,\zeta_m,z,\bar z)=0\,,\nonumber\\ &&\label{conformdiffeqN}
\end{eqnarray}
the corresponding spin average satisfies \cite{level2}
\begin{eqnarray}
&&\left[-{4\over 3}\,{\partial^2\over\partial z^2}+{1\over z-\bar z}\,{\partial\over\partial\bar z}+{1/16\over(z-\bar z)^2} +\sum_{j=1}^m{1\over z-\zeta_j}\,{\partial\over\partial \zeta_j}+\langle T(z)\rangle_{+f+f+\,\dots}\right]\langle\sigma\rangle_{+f+f+\,\dots}=0\,.\nonumber\\ &&\label{conformdiffeqsig}
\end{eqnarray}
Combining Eqs.~(\ref{NoverD2}), (\ref{conformdiffeqN}), and (\ref{conformdiffeqsig}) leads to 
\begin{equation}
\langle T(z)\rangle_{+f+f+\,\dots}=\sum_{j=1}^m\left[{1/16\over(z-\zeta_j)^2}+{1\over z-\zeta_j}\,{\partial\over\partial \zeta_j}\,\ln G_\sigma^{(m,1)}(\zeta_1,\dots,\zeta_m)\right]\,.\label{T+f+etc}
\end{equation}
A similar calculation based on the differential equations for any of the correlation functions $\langle\sigma_1\dots\sigma_\ell\,\epsilon_{\ell+1}\dots\epsilon_n\rangle$ with $+f+f+\dots$ boundary conditions leads to exactly the same stress tensor, since, for each of these correlation functions the denominator $D$ in the in the $N/D$ form, is also proportional to $G_\sigma^{(m,1)}(\zeta_1,\dots,\zeta_m)$. 
 
In the case of $+f+$ boundary conditions, corresponding to $m=2$, combining Eqs.~(\ref{G2sigma}) and (\ref{T+f+etc}) yields the same average stress tensor as in Eq.~(\ref{Taba}), with $t_{ab}=t_{ba}=t_{f+}={1\over 16}$.
 
If there are more than two points $\zeta_1$, $\zeta_2$ on the $x$ axis at which the boundary condition changes, the explicit form of average stress tensor is no longer determined by the elementary considerations that imply Eq.~(\ref{Taba}), but follows from conformal-invariance theory. For $+f+f+$ boundary conditions or $m=4$, Eqs.~(\ref{G41sigma}) and (\ref{T+f+etc}) lead to  
\begin{eqnarray}
 \langle T\rangle_{+f+f+}
 &=&{\textstyle{1\over 16}}\left({1\over z-\zeta_1}-{1\over z-\zeta_2}\right)^2+{\textstyle{1\over 16}}\left({1\over z-\zeta_3}-{1\over z-\zeta_4}\right)^2\nonumber\\
 &&+{1\over 8}{\sqrt{\zeta_{31}\zeta_{42}} -\sqrt{\zeta_{41}\zeta_{32}}\over\sqrt{\zeta_{31}\zeta_{42}} +\sqrt{\zeta_{41}\zeta_{32}}} \left({1\over z-\zeta_1}-{1\over z-\zeta_2}\right)\left({1\over z-\zeta_3}-{1\over z-\zeta_4}\right)\,,\nonumber\\ \label{T+f+f+}
\end{eqnarray} 
where $\zeta_{ij}=\zeta_i-\zeta_j$																			

Now we derive a formula analogous to Eq.~(\ref{T+f+etc}) for the semi-infinite critical Ising model with the alternating boundary condition $+-+-+\,\dots$. The correlation functions of $\sigma$ and $\epsilon$ in this system are analyzed in \cite{TWBG2}. In particular,\begin{equation}\langle\epsilon_1\dots\epsilon_n\rangle_{+-+-+\,\dots}=i^n\,{G_\epsilon^{(m+2n)}(\zeta_1,\dots,\zeta_m,z_1,\bar z_1,\dots,z_n,\bar z_n)\over G_\epsilon^{(m)}(\zeta_1,\dots,\zeta_m)}\,,
\end{equation}
where the function $G_\epsilon^{(n)}(z_1,\dots,z_n)$ is defined in Eq.~(\ref{Gneps}). Recalling  that the scaling index for the energy is $\Delta_\epsilon={1\over 2}$ and carrying out a calculation similar to the one leading to Eq.~(\ref{T+f+etc}) leads to
\begin{equation}
\langle T(z)\rangle_{+-+-+\,\dots}=\sum_{j=1}^m\left[{1/2\over(z-\zeta_j)^2}+{1\over z-\zeta_j}\,{\partial\over\partial \zeta_j}\,\ln G_\epsilon^{(m)}(\zeta_1,\dots,\zeta_m)\right]\,\label{T+-+etc}
\end{equation}
where 
\begin{equation}
G_\epsilon^{(n)}(\zeta_1,\dots,\zeta_n)={\rm Pf}^{(n)}{1\over\zeta_{ij}}\,.\label{T+-+etcPf}
\end{equation}
Equations (\ref{T+-+etc}) and (\ref{T+-+etcPf}) are consistent with Eq. (D4) in Ref.~\cite{EETWB} but have a simpler form

In the case of  $+-+$ boundary conditions, corresponding to $m=2$, combining Eqs.~(\ref{G2eps}) and (\ref{T+-+etc}) yields the same average stress tensor as in Eq.~(\ref{Taba}), with $t_{ab}={1\over 2}$. For $+-+-+$ boundary conditions or $m=4$, Eqs.~(\ref{G4eps}) and (\ref{T+-+etc}) imply 
\begin{eqnarray}
 &&\langle T\rangle_{+-+-+}
 =\textstyle{1/2\over(z-\zeta_1)^2}+{1/2\over(z-\zeta_2)^2}+{1/2\over(z-\zeta_3)^2}+{1/2\over(z-\zeta_4)^2}\nonumber\\[2mm]
 &&\qquad -\textstyle\Big\{\textstyle\left[{1\over (z-\zeta_1)(z-\zeta_2)}+{1\over (z-\zeta_3)(z-\zeta_4)}\right]{1\over \zeta_{12}\zeta_{34}}
 -\textstyle\left[{1\over (z-\zeta_1)(z-\zeta_3)}+{1\over (z-\zeta_2)(z-\zeta_4)}\right]{1\over \zeta_{13}\zeta_{24}}\nonumber\\[2mm]
 &&\qquad +\textstyle\left[{1\over (z-\zeta_1)(z-\zeta_4)}+{1\over (z-\zeta_2)(z-\zeta_3)}\right]{1\over \zeta_{14}\zeta_{23}}\Big\}\Big/
 \left({1\over\zeta_{12}\zeta_{34}}-{1\over\zeta_{13}\zeta_{24}}+{1\over\zeta_{14}\zeta_{23}}\right)\,,\label{T+-+-+}
\end{eqnarray} 
in agreement with Eq.~(D3) in Ref.~\cite{EETWB}.

\subsection{Boundary condition $-f+$}
\label{mfp}
In this section we consider the $n$-spin correlation function $\langle\sigma_1\dots\sigma_n\rangle_{-f+}$ of the semi-infinite Ising model with spin-down boundary conditions on the $x$ axis for $x<\zeta_1$, free spins for $\zeta_1<x<\zeta_2$,  and spin-up for $x>\zeta_2$, respectively.  For reasons that will become clear, it is  convenient to begin, not with $-f+$, but with the $f+-f$ boundary corresponding to free spins for $x<\zeta_a$, spin up for $\zeta_a<x<\zeta_b$, spin down for $\zeta_b<x<\zeta_c$, and free spins for $z>\zeta_c$. Once $\langle\sigma_1\dots\sigma_n\rangle_{f+-f}$ has been determined, it is simple to obtain $\langle\sigma_1\dots\sigma_n\rangle_{-f+}$ with a conformal coordinate transformation involving inversion about an appropriate point on the boundary.

Recall that the amplitudes  $t_{f+}=t_{f-}={1\over 16}$ and $t_{+-}={1\over 2}$, introduced below Eq.~(\ref{Tab}), equal the scaling indices of $\sigma$ and $\epsilon$, respectively. Accordingly, $\langle\sigma_1\dots\sigma_n\rangle_{f+-f}$ in the variables $(\zeta_a,\zeta_b,\zeta_c,z_1,\bar z_1,\dots,z_n,\bar z_n)$                   is determined by the same conformal differential equations as the bulk correlation function $\langle\sigma_a\epsilon_b\sigma_c\sigma_1\dots\sigma_{2n}\rangle$ in the variables $(z_a,z_b,z_c,z_1,z_2,\dots,z_{2n})$. One possible  strategy for calculating $\langle\sigma_1\dots\sigma_n\rangle_{f+-f}$ is to attempt to solve these differential equations, using the approach of \cite{TWBG1,TWBX}. 

Here we follow a different strategy. Setting $\zeta_a=-\zeta$, $\zeta_b=0$, and $\zeta_c=\zeta$, we switch the boundary condition at the origin with the help of two disorder operators \cite{KC,Cardydisorderop,TWBG1}. The advantage of this approach is that both the spin operator $\sigma$ and the dual disorder operator $\mu$ have bulk scaling dimension $\Delta={1\over 16}$, and the solutions of the relevant conformal different equations are the known functions $G^{(n,\alpha)}(z_1,\dots,z_n)$ in Eq.~(\ref{Gnsigma}).  

Following \cite{KC,Cardydisorderop,TWBG1}, we express the desired correlation function as 
\begin{equation}
\langle\sigma_1\dots\sigma_n\rangle_{f+-f}=\lim_{\substack{Y_1\to 0 \\ Y_2\to\infty}}{\langle\mu(iY_1,-iY_1)\mu(iY_2,-iY_2)\sigma(z_1,\bar z_1)\dots\sigma(z_n,\bar z_n)\rangle_{f+f}\over\langle\mu(iY_1,-iY_1)\mu(iY_2,-iY_2)\rangle_{f+f}}\,,\label{sigdotssigf+-f1}
\end{equation}
in terms of correlation functions with the $f+f$ boundary condition, with free spins for $x<\zeta$ and $x>\zeta$ and spin up for $-\zeta<x<\zeta$. In the indicated limit the two disorder operators $\mu$ introduce a ladder of antiferrogmetic bonds along the positive $y$ axis, leading from $f+f$ to $f+-f$ boundary conditions. 

Writing  both the correlation functions in the numerator and denominator in Eq.~(\ref{sigdotssigf+-f1}) in $N/D$ form,
 as in Eq.~(\ref{NoverD1}) leads to
 \begin{equation}
\langle\sigma_1\dots\sigma_n\rangle_{f+-f}=\lim_{\substack{Y_1\to 0 \\ Y_2\to\infty}}{N_1(iY_1,-iY_1,iY_2,-iY_2,-\zeta,\zeta,z_1,\bar z_1,\dots,z_n,\bar z_n)
\over N_2(iY_1,-iY_1,iY_2,-iY_2,-\zeta,\zeta)} \,,\label{sigdotssigf+-f2}
\end{equation}
Since the spin operator $\sigma$, the dual operator $\mu$, and the $f+$ and $+f$ boundary operators all have scaling index $\Delta_\sigma=\Delta_\mu=t_{f+}=t_{+f}={1\over 16}$, the function $N_1$ in Eq.~(\ref{sigdotssigf+-f2}) satisfies the same differential equations in its $2n+6$ arguments as the bulk correlation function $\langle\sigma\dots\sigma_{2n+6}\rangle_{\rm bulk}$ in the variables $z_1,\dots,z_{2n+6}$. Thus, $N_1$ is an appropriate linear combination of the $2^{n+2}$ functions $G_\sigma^{(2n+6,\alpha)}(iY_1,-iY_1, iY_2,-iY_2,-\zeta,\zeta,z_1,\bar z_1,\dots,z_n,\bar z_n)$ defined in Eq.~(\ref{Gneps}). Similarly, $N_2$ is a linear combination of the 4 functions $G_\sigma^{(6,\alpha)}(iY_1,-iY_1, iY_2,-iY_2,-\zeta,\zeta)$.

\subsubsection{One-point function $\langle\sigma\rangle_{-f+}$}
In the special case $n=1$, the function $N_1$ in Eq.~(\ref{sigdotssigf+-f2}) is an appropriate linear combination of the 8 functions $G_\sigma^{(8,\alpha)}$  defined by Eqs.~(\ref{Gnsigma}) and (\ref{xidef}), with 
\begin{eqnarray}
&&\xi_{13}=\left({Y_2-Y_1\over Y_2+Y_1}\right)^{1/2},\quad\xi_{15}=\left[{Y_1-i\zeta\over(Y_1^2+\zeta^2)^{1/2}}\right]^{1/2}\,,\quad \xi_{17}=\left[{x^2+(y-Y_1)^2\over 
x^2+(y+Y_1)^2}\right]^{1/4},\nonumber\\
&&\xi_{35}=\left[{Y_2-i\zeta\over(Y_2^2+\zeta^2)^{1/2}}\right]^{1/2}\,,\quad \xi_{37}=\left[{x^2+(y-Y_2)^2\over 
x^2+(y+Y_2)^2}\right]^{1/4},\quad\xi_{57}=e^{-i(\theta_2-\theta_1)/2}\,,\nonumber\\
&&e^{\theta_1}={z+\zeta\over|z+\zeta|}={x+\zeta+iy \over\sqrt{(x+\zeta)^2+y^2}}\,,\quad e^{i\theta_2}={z-\zeta\over|z-\zeta|}={x-\zeta+iy\over\sqrt{(x-\zeta)^2+y^2}}\,.\label{defsmfp}
\end{eqnarray}
Examining the leading asymptotic contribution of each of the  $G_\sigma^{(8,\alpha)}$ for both $Y_1$ and $1/Y_2$ small, we find that only the contribution of  $G_\sigma^{(8,7)}$ is consistent with the expected sign change in $\langle\sigma\rangle_{f+-f}$ as $x$ changes sign and the expected asymptotic behavior for the $f+-f$ boundary condition as $y\to 0$. Choosing the proportionality constant for consistency with Eq.~(\ref{sigfixed}), with the plus sign for $-\zeta<x<0$ and the minus sign for $0<x<\zeta$, we find
\begin{eqnarray}
&&\langle\sigma\rangle_{f+-f}=-\left({2\over y}\right)^{1/8}\sqrt{\sin\left(\textstyle{1\over 2}{\theta_2}-\textstyle{1\over 2}{\theta_1}\right)-{y\zeta\over x^2+y^2}\cos\left(\textstyle{1\over 2}\theta_2-\textstyle{1\over 2}\theta_1\right)}\,.\label{sigminusfreeplustrig0}
\end{eqnarray}

To obtain $\langle\sigma\rangle_{-f+}$ for the desired boundary condition of down spins for $x<\zeta_1$, free spins for $\zeta_1<x<\zeta_2$, and up spins for $z>\zeta_2$, use of the conformal transformation property
\begin{equation}
 \langle\sigma(z',\bar z')\rangle_{-f+}=\left\vert{dz'\over dz}\right\vert^{-1/8}\langle\sigma(z,\bar z)\rangle_{f+-f}\label{conformtrans}
 \end{equation}
 together with the mapping 
\begin{equation}
z'={\textstyle{1\over 2}}(\zeta_1+{\textstyle\zeta_2)+{1\over 2}}(\zeta_1-\zeta_2){\zeta\over z}\label{confomap}
\end{equation}
to change the boundary geometry. This leads to 
 \begin{equation}
\langle\sigma\rangle_{-f+}=\left({2\over y}\right)^{1/8}\sqrt{\cos\left(\textstyle{1\over 2}\gamma_{1,1}\right)-{2y\over\zeta_2-\zeta_1}\sin\left(\textstyle{1\over 2}\gamma_{1,1}\right)}\,,\label{sigmfp0}
\end{equation}
the main result of this subsection, where we have dropped the primes for simplicity. The quantity $\gamma_{1,1}$ in Eq.~(\ref{sigmfp0}) is the same as in Eq.~(\ref{sigpfp}) and defined by Eqs.~(\ref{gammakelldef1}) and (\ref{gammakelldef2}). 

On making use of Eqs.~(\ref{cosgammakellover2})-(\ref{singammakell}), Eq.~(\ref{sigmfp0}) can be expressed entirely in terms of the Cartesian coordinates $(x,y)$. 
The one-point function (\ref{sigmfp}) is expected to vanish on the half line $x={1\over 2}(\zeta_1+\zeta_2)$, $y>0$, since all points on this half line are equidistant from the up and down pointing boundary spins. Defining $\Delta x\equiv x-{1\over 2}(\zeta_1+\zeta_2)$, we choose 
the square root in Eq.~(\ref{sigmfp0}) to be positive for $\Delta x>0$ and negative for $\Delta x<0$, so that  $\langle\sigma\rangle_{-f+}$ is an odd function of $\Delta x$. Expanding the argument of the square root in powers of $\Delta x$, one finds 
\begin{equation}
\langle\sigma\rangle_{-f+}=\left({2\over y}\right)^{-1/8}\sqrt{{y\over\left({1\over 4}\,\zeta_{21}^2+y^2\right)^{3/2}}\,\Delta x^2+{\rm O}\left(\Delta x^4\right)}\,,
\end{equation}
where $\zeta_{21}\equiv\zeta_2-\zeta_1$,
consistent with a smooth, analytic continuation between the positive and negative branches at $\Delta x=0$.

\subsubsection{One and two-point averages for $-f+$ and $f+-$ boundary conditions.}
\paragraph{$-f+$ boundary conditions}
 
 To calculate the spin-spin correlation function $\langle\sigma_1\sigma_2\rangle_{-f+}$,  we again begin with $f+-f$ boundary conditions and with  Eq.~(\ref{sigdotssigf+-f2}) for $n=2$. Examining the leading asymptotic contribution of each of the 16 functions $G_\sigma^{(10,\alpha)}(iY_1,-iY_1,iY_2,-iY_2,-\zeta,\zeta,z_1,\bar z_1,z_2,\bar z_2)$ defined by Eq.~(\ref{Gnsigma}), for both $Y_1$ and $1/Y_2$ small, we find that only $G_\sigma^{(10,11)}$ yields an expression for $\langle\sigma_1\sigma_2\rangle_{-f+}$ consistent with the operator product expansion (\ref{OPEsigsig}) for small $\vert z_1-z_2\vert$ and the expected asymptotic behavior in the limits such as $x_1\to \pm\infty$, $y_1\to 0$, $y_2\to 0$. Transforming from $f+-f$ to the $-f+$ geometry, as in the preceding subsection, leads to the result for  $\langle\sigma_1\sigma_2\rangle_{-f+}$ in Eq.~(\ref{sigsigmfp}).  Comparing the result with the operator product expansion (\ref{OPEsigsig}) leads to the expression for  $\langle\epsilon_1\rangle_{-f+}$ in Eq.~(\ref{epsmfp}).
 
 Proceeding in the same way, we have constructed $\langle\sigma_1\dots\sigma_n\rangle_{-f+}$ for $n=3$ and 4, beginning with  Eq.~(\ref{sigdotssigf+-f2}) and the families of 32 functions $G_\sigma^{(12,\alpha)}$ and 64 functions $G_\sigma^{(14,\alpha)}$, respectively. Comparing the results with the operator product expansion (\ref{OPEsigsig}) leads to expressions (\ref{sigepsmfp}) for  $\langle\sigma_1\epsilon_2\rangle_{-f+}$ and (\ref{epsepsmfp}) for $\langle\epsilon_1\epsilon_2\rangle_{-f+}$.
 
In terms of the variables $\rho$, $\gamma_{k,\ell}$, and $\chi$ defined in Eqs.~(\ref{rhodef2}), (\ref{gammakelldef1}), and (\ref{chidef}),
\begin{eqnarray}
&&\langle\sigma_1\rangle_{-f+}=\left({2\over y_1}\right)^{1/8}\sqrt{\cos\left(\textstyle{1\over 2}\gamma_{1,1}\right)-{2y_1\over\zeta_2-\zeta_1}\sin\left(\textstyle{1\over 2}\gamma_{1,1}\right)}\,,\label{sigmfp}\\[3mm]
&& \langle\epsilon_1\rangle_{-f+}=-{1\over 2y_1}\left(\cos\gamma_{1,1}-{4y_1\over\zeta_2-\zeta_1}\sin\gamma_{1,1}\right)\,,\label{epsmfp}
\end{eqnarray}

\begin{eqnarray}
&&\langle\sigma_1\sigma_2\rangle_{-f+}=\left({1\over 4y_1y_2}\right)^{1/8}\Bigg[{1\over\sqrt{\rho}}\cos\left(\textstyle{1\over 2}\gamma_{1,1}-\textstyle{1\over 2}\gamma_{2,1}\right)+\sqrt{\rho}\cos\left(\textstyle{1\over 2}\gamma_{1,1}+\textstyle{1\over 2}\gamma_{2,1}\right)\nonumber\\  
&&\qquad-{2\over\sqrt{\rho}}\,{y_1-y_2\over\zeta_2-\zeta_1}\sin\left(\textstyle{1\over 2}\gamma_{1,1}-\textstyle{1\over 2}\gamma_{2,1}\right)-2\sqrt{\rho}\;{y_1+y_2\over\zeta_2-\zeta_1}\sin\left(\textstyle{1\over 2}\gamma_{1,1}+\textstyle{1\over 2}\gamma_{2,1}\right)\Bigg]^{1/2},\label{sigsigmfp}
\end{eqnarray}

\begin{eqnarray}
&&\langle\sigma_1\epsilon_2\rangle_{-f+}=-
{1\over 2}\left({2\over y_1}\right)^{1/8}\left({1\over 2y_2}\right)\,\Bigg[{1\over\rho}\cos\left(\textstyle{1\over 2}\gamma_{1,1}-\gamma_{2,1}\right)
\nonumber\\[2mm] &&\qquad+\rho\cos\left(\textstyle{1\over 2}\gamma_{1,1}+\gamma_{2,1}\right)-
{2\over\rho}\,{y_1-2y_2\over\zeta_2-\zeta_1}\,\sin\left(\textstyle{1\over 2}\gamma_{1,1}-\gamma_{2,1}\right)\nonumber\\[2mm] 
 &&\qquad -2\rho\,{y_1+2y_2\over\zeta_2-\zeta_1}\,\sin\left(\textstyle{1\over 2}\gamma_{1,1}+\gamma_{2,1}\right)\Bigg] \Big/ \sqrt{\cos\left(\textstyle{1\over 2}\gamma_{1,1}\right)-{2y_1\over\zeta_2-\zeta_1}\,\sin\left(\textstyle{1\over 2}\gamma_{1,1}\right)}\,,\label{sigepsmfp}
 \end{eqnarray}
 
\begin{eqnarray}
&&\langle\epsilon_1\epsilon_2\rangle_{-f+}=-{1\over 8y_1y_2}\Bigg[\left(1-{2\over\rho^2}+{16y_1y_2\over(\zeta_2-\zeta_1)^2}\right)\cos(\gamma_{1,1}-\gamma_{2,1})\nonumber\\[2mm]
&&\qquad +\left(1-2\rho^2-{16y_1y_2\over(\zeta_2-\zeta_1)^2}\right)\cos(\gamma_{1,1}+\gamma_{2,1})-4\left(1-{2\over\rho^2}\right)\,{y_1-y_2\over\zeta_2-\zeta_1}\sin(\gamma_{1,1}-\gamma_{2,1})\nonumber\\[2mm]
&&\qquad -4\left(1-2\rho^2\right)\,{y_1+y_2\over\zeta_2-\zeta_1}\sin(\gamma_{1,1}+\gamma_{2,1})\Bigg]\,.\label{epsepsmfp}
\end{eqnarray}

In Fig.~\ref{fig1} the one-point averages $\langle\sigma\rangle_{-f+}$ and $\langle\epsilon\rangle_{-f+}$ in Eqs. (\ref{sigmfp}) and (\ref{epsmfp}) are plotted as functions of $x$ for $y={1\over 4}$ and $\zeta_1=-\zeta_2=-1$. The quantities $\langle\sigma\rangle_{+f+}$ and $\langle\epsilon\rangle_{+f+}$ in Eqs. (\ref{sigpfp}) and (\ref{epspfp}) are shown for comparison. The curves for $\langle\sigma\rangle_{-f+}$ and $\langle\sigma\rangle_{+f+}$ look qualitatively as expected, reflecting the odd and even dependence on $x$, respectively, and approaching 
$\langle\sigma\rangle_+$ or $\langle\sigma\rangle_-$ for $|x|\to\infty$. 

Since the $-f+$ boundary condition is less conducive to ordering than the  $+f+$ boundary condition, the curve for $\langle\epsilon\rangle_{-f+}$ in Fig.~\ref{fig1} lies above the curve for
$\langle\epsilon\rangle_{+f+}$.  For sufficiently small $|x|$, it even rises above the dashed line representing $\langle\epsilon\rangle_f$.

Setting  $\zeta_1=-\zeta_2=-1$ in Eqs.~(\ref{epsfixedfree}) and (\ref{epsmfp}), we find that $\langle\epsilon\rangle_{-f+}$  exceeds 
$\langle\epsilon\rangle_f$ for $|x|<({1\over 2}+y^2)^{1/2}$ and has a maximum at $x=0$ with height ratio $\langle\epsilon\rangle_{-f+}/\langle\epsilon\rangle_f=(1+3y^2)/(1+y^2)$. Thus, for $y={1\over 4}$, as in Fig.~\ref{fig1}, the corresponding interval and height ratio are $|x|<{3\over 4}$ and ${19\over 17}$, respectively. For $y\gg 1$, the expressions for the interval and height ratio yield $|x|<y$ and 3. These results are easily checked by noting that $\langle\epsilon\rangle_{-f+}\to\langle\epsilon\rangle_{-+}$ for $y\gg \zeta_2-\zeta_1$  and using the explicit form of $\langle\epsilon\rangle_{-+}$ in Eq.~(\ref{sigandepsab}) .\\

\paragraph{$f+-$ boundary conditions}
The one and two-point averages for $-f+$ boundary conditions in Eqs.~(\ref{sigmfp}) and (\ref{epsepsmfp}) can be transformed into results for $f+-$ boundaries using the conformal mapping
\begin{equation}
z'=-{(\zeta_2-\zeta_1)(\zeta_2'-\zeta_1')\over z-\zeta_1}+\zeta_2'\,,\quad z=-{(\zeta_2-\zeta_1)(\zeta_2'-\zeta_1')\over z'-\zeta_2'}+\zeta_1\,,\label{mfptofpm}
\end{equation}
which maps $z=\zeta_1,\,\zeta_2,\,\infty$ on to $z'=\infty,\,\zeta_1',\,\zeta_2'$, respectively. 
In terms of the primed variables, 
\begin{eqnarray}
&&\gamma_{1,1}={\rm arg}\left({z_1-\zeta_2\over z-\zeta_1}\right)={\rm arg}\left({z_1'-\zeta_1'\over\zeta_2'-\zeta_1'}\right)={\rm arg}\left({|z_1'-\zeta_1'|e^{i\vartheta_1'}\over\zeta_2'-\zeta_1'}\right)=\vartheta_1'\,,\label{gamma11prime}\\[2mm]
&&{y_1\over \zeta_2-\zeta_1}={(\zeta_2'-\zeta_1')y_1'\over\left\vert z'-\zeta_2'\right\vert^2}\,,\label{yprime}\end{eqnarray}
where we have used the definition (\ref{gammakelldef2}) of $\gamma_{1,1}$. Beginning with Eqs.~(\ref{sigmfp}) -(\ref{epsepsmfp}), using 
Eqs.~(\ref{gamma11prime}) and (\ref{yprime}) and the transformation property analogous to (\ref{conformtrans}), and dropping primes in the final expression, we obtain
\begin{eqnarray}
&&\langle\sigma_1\rangle_{f+-}=\left({2\over y_1}\right)^{1/8}\sqrt{\cos{\textstyle\vartheta_1\over 2}-\,{2\,\zeta_{21}y_1\over\left\vert z-\zeta_2\right\vert^2}\sin{\textstyle\vartheta_1\over  2}}\label{sigfpm}\\[3mm]
&& \langle\epsilon_1\rangle_{f+-}=-{1\over 2y_1}\left[\cos\vartheta_1-{4\,\zeta_{21}y_1\over\left\vert z-\zeta_2\right\vert^2}\sin\vartheta_1\right]\,,\label{epsfpm}
\end{eqnarray}
where $\vartheta_1={\rm arg}(z_1-\zeta_1)$, and corresponding results for the two-point averages.

\subsubsection{Average stress tensor $\langle T(z)\rangle_{-f+}$} 
 
The average stress tensor $\langle T(z)\rangle_{_f+}$ is given by Eq.~(\ref{Tabc}) with $t_{ab}=t_{bc}=t_{f+}={1\over 16}$ and $t_{ac}=t_{-+}={1\over 2}$, which leads to
\begin{equation}
 \langle T(z)\rangle_{-f+}={1/16\over (z-\zeta_1)^2}+{1/16\over (z-\zeta_2)^2}+{3/8\over(z-\zeta_1)(z-\zeta_2)}\,.\label{Tmfp}
 \end{equation}
 
 According to the conformal theory the one-point averages of $\sigma$ and $\epsilon$ for $-f+$ boundary conditions satisfy \cite{level2}
\begin{eqnarray}
 &&\bigg( -{4\over 3}\,{\partial^2\over \partial z^2}+ {1\over z-\bar z}\,{\partial\over \partial\bar z}+{1/16\over (z-\bar z)^2}\nonumber\\ 
 &&\qquad\qquad\qquad +{1\over z-\zeta_1}\,{\partial\over \partial\zeta_1}+ {1\over z-\zeta_2}\,{\partial\over \partial\zeta_2}+\langle T(z)\rangle_{-f+}\bigg)\langle\sigma\rangle_{-f+}=0\,,\label{diffeqsigmfp1}\\
 &&\bigg( -{3\over 4}\,{\partial^2\over \partial z^2}+ {1\over z-\bar z}\,{\partial\over \partial\bar z}+{1/2\over (z-\bar z)^2}\nonumber\\ 
 &&\qquad\qquad\qquad +{1\over z-\zeta_1}\,{\partial\over \partial\zeta_1}+ {1\over z-\zeta_2}\,{\partial\over \partial\zeta_2}+\langle T(z)\rangle_{-f+}\bigg)\langle\epsilon\rangle_{-f+}=0\label{diffeqepsmfp1}
 \end{eqnarray}
As a check on our results (\ref{sigmfp}) and (\ref{epsmfp}) for the one-point functions, we have confirmed that substituting them into Eqs.~(\ref{diffeqsigmfp1}) and (\ref{diffeqsigmfp1}) and solving for $\langle T(z)\rangle_{-f+}$ reproduces the average stress tensor in Eq.~(\ref{Tmfp}). The two-point functions  $\langle\sigma_1\sigma_2\rangle_{-f+}$, $\langle\sigma_1\epsilon_2\rangle_{-f+}$, and $\langle\epsilon_1\epsilon_2\rangle_{-f+}$ satisfy differential equations which are obvious generalizations of Eqs.~(\ref{diffeqsigmfp1}) and (\ref{diffeqepsmfp1}). Here also our results (\ref{sigsigmfp}),  (\ref{sigepsmfp}), and (\ref{epsepsmfp}) and the differential equations lead to  the average stress tensor (\ref{Tmfp}). 

\subsection{Casimir interaction of a wedge with the boundary}
\label{wedge}
Consider a wedge-shaped inclusion pointing perpendicularly toward the $x$ axis in a critical Ising system defined on the upper half $z$ plane . The edges of the wedge form angles $\alpha$ and $\pi - \alpha$, where $0< \alpha < \pi /2$, with the $x$ axis and intersect at the tip of the wedge, which is on the $y$ axis a distance $D$ from the origin. This roughly resembles the geometry of an atomic force microscope. 

To calculate the Casimir force acting on the wedge, we proceed as in Ref. \cite{EETWB} and use the conformal transformation $z(w)$ with derivative
\begin{eqnarray} \label{wedgetohalf}
{dz \over dw}= - {D \over E(\alpha)} \; e^{-i\alpha} w^{-(1+\alpha/\pi)}(w-1)^{2\alpha/\pi} \, , \quad
E(\alpha)= 2 \int_{0}^{\pi/2} d\psi (2 \sin \psi)^{2\alpha/\pi} \,,
\end{eqnarray}
to map the empty upper half $w=u+iv$ plane onto the simply-connected region of the $z=x+iy$ plane between the wedge and the $x$ axis.   
Under this transformation the segments $-\infty < u < 0$, $0<u<1$, and $1<u$ of the $u$ axis map onto the $x$ axis X, the right boundary WR, and the left boundary WL of the wedge, respectively. According to Ref. \cite{EETWB} the wedge experiences the force 
\begin{eqnarray} \label{force} 
&&(f_x, \, f_y)/(k_B T) = -({\rm Im}, \, {\rm Re}) \left(\tau^{(T)}+\tau^{(S)} \right)\,,\label{force1}\\[2mm]
&&\left[ \tau^{(T)},\,\tau^{(S)} \right]={1\over\pi}\int_Cdw \, {1 \over z'(w)} \,\left[ \langle
T(w) \rangle \, , \, - {1 \over 24}\, \{z,w\} \right] \, ,\label{tautau}
\end{eqnarray}
where
the integration path $C$ is along the $u$ axis from $w=0$ to $+\infty$ and  passes above the singularity at $w=1$. The quantity $\langle T(w) \rangle$ in Eq.~(\ref{tautau}) is the average stress tensor in the empty upper half $w$ plane, and $\{z,w\} \equiv z'''(w)/z'(w) - (3/2) \left[z''(w)/z'(w)\right]^2$ is the Schwarzian derivative, which equals
\begin{eqnarray} \label{wedgeS}
\{z,w\} = \left( 1+{\alpha \over \pi} \right) \left[ \left( 1- {\alpha \over \pi}\right)  {1 \over 2 w^2}  -  {2\alpha \over \pi}  {1 \over w(w-1)^2} \right] 
\end{eqnarray}
for the mapping (\ref{wedgetohalf}). 
Unlike $\{z,w\}$, $\langle T(w) \rangle$ depends on the boundary conditions in the wedge geometry, since they determine the boundary conditions on the corresponding three segments of the $u$ axis. 

We now examine the Casimir force in detail for the boundary conditions $f$, $+$, and $-$, on X, WR, and WL, respectively. This is an especially interesting case, since the Casimir force on the wedge reverses direction at a critical value of the apex angle, as we shall see.
According to Eq. (\ref{Tabc}), with $z$ replaced by $w$, 
\begin{eqnarray} \label{wedgeS}
\langle T(w) \rangle_{f+-}^{(\zeta_1 =0, \, \zeta_2 =1)} =  {1/16 \over w^2} + {1/2 \over  w(w-1)^2}  \, .
\end{eqnarray}
 Substituting this expression for $T(w)$ in Eq. (\ref{tautau}), and using the relation
\begin{eqnarray} \label{integ}
\int\limits_{0}^{+\infty} du (u-a+i0)^{-\nu} \, u^{\mu-1} \, = \, a^{\mu-\nu} i^{2(\mu-\nu)} B(\mu,\nu-\mu) \, , \quad a>0 \, ,
\end{eqnarray}
where $B$ is the beta function, corresponding to formula 3.194.3 in Ref. \cite{G+R}, we obtain 
\begin{eqnarray} \label{tau,tau}
&&\left[\tau^{(S)} \, , \; \tau^{(T)} \right] = - {1 \over D} \, E(\alpha) \, G(\alpha) \left[ -\left(1+{\alpha \over \pi}  \right)^2 \, , \;   3\left( 1-2{\alpha \over \pi} \right) \right] \,,\\ 
&&G(\alpha)={\Gamma^2 (\alpha/\pi) \over 48 \pi \Gamma(2\alpha/\pi)} \, {1 \over 1+(2\alpha/\pi)}\,.
\end{eqnarray}
Together with Eq. (\ref{force}), this implies $f_x =0$ and 
\begin{eqnarray} \label{force2}
{f_y  \over k_B T}  =  {1 \over D} \, E(\alpha) \, G(\alpha) \, \left[ 2-8{\alpha \over \pi} - \left({\alpha \over \pi} \right)^2 \,  \right] \, .
\end{eqnarray}

Rewriting the square bracket in Eq.(\ref{force2}) as
\begin{eqnarray} \label{alpha0}
[ \;]=(\alpha_0 -\alpha)(\alpha_0 +\alpha +8 \pi)/\pi^2 \, , \quad \alpha_0 = (3 \sqrt{2} -4) \pi = 0.243 \pi =43.7^{\,\circ}\, ,
\end{eqnarray}
and noting that $E(\alpha)$ and $G(\alpha)$ are positive, we find that
$f_y$ is positive for $0<\alpha<\alpha_0$ and negative for $\alpha_0<\alpha<\pi/2 $, corresponding to repulsion and attraction, respectively,  of the wedge by the boundary. In terms of the apex angle $\beta=\pi-2\alpha$, the force is attractive for $0<\beta<\beta_0$ and repulsive for $\beta_0<\beta<\pi$, where $\beta_0=(9-6\sqrt{2})\pi=92.6^{\,\circ}$.

This behavior is consistent with the following picture: For small $\beta$, the wedge is almost a needle, and the dominant force is between its tip and the $f$ boundary. Since the junction of the $+$ and $-$ boundaries at the tip and the $f$ boundary both favor disorder, the overall force is attractive. For $\beta$ near $\pi$, on the other hand, the $+$ and $-$ boundaries of the wedge lie along the positive and negative $x$ axes, respectively, both of which have boundary condition $f$. Since the $f$ boundary repels both $+$ and $-$ boundaries, the overall force on the wedge is repulsive.

In the limit of a $-+$ needle, $\alpha = \pi /2$, $\tau^{(T)}=0$, $E=4$, $G= 1/96$, and $f_y/(k_B T)=-3/(32 D)$. This $f_y$  is the same as for an $aa$ needle in the upper half $z$ plane with a uniform $a$ boundary along the $x$ axis \cite{EETWB}. In the latter case the empty upper half $w$ plane also has uniform boundary condition $a$, so that $\langle T(w) \rangle$ vanishes.

\section{Boundary-operator expansions in systems\\ with mixed boundary conditions}\label{secMBOE}
\subsection{Boundary-operator expansion away from switching points} \label{away}
Boundary-operator expansions have been studied extensively in semi-infinite critical systems with uniform boundary conditions \cite{Diehl,CardyLewellen,EEStap}. In the expansion of a primary operator $\phi(x,y)$, with a distance $y$ from the boundary much smaller than the other lengths that characterize the system, $\phi(x,y)$ is expressed as a series of $y$-independent boundary operators with increasing scaling dimension, multiplied by appropriate powers of $y$. For the Ising model defined on the upper half plane with uniform boundary condition $h$ on the $x$ axis and for the pairs 
$(\phi, h)=(\sigma,+),\,(\sigma,-),\,(\epsilon,+),\,(\epsilon,-),\,(\epsilon,f)$, the leading boundary operator is the stress tensor $T(z)$ evaluated on the $x$ axis. To lowest order the expansion reads
\begin{eqnarray} 
\phi(x, y)-\langle \phi \rangle_h \to \mu_h^{(\phi)}\, y^{2-x_{\phi}} T(x)  \, , \quad  y \to 0 \, , \label{BOE}
\end{eqnarray}
where $x_\phi=2\Delta_\phi$ is the scaling dimension of $\phi$.
The averages $\langle \phi \rangle_h$ in Eq. (\ref{BOE}) for $\phi$ equal to $\sigma$ and $\epsilon$ are given in Eqs.~(\ref{sigfixed}) and (\ref{epsfixedfree}), and the universal amplitudes $\mu_h^{(\phi)}$ are
\begin{eqnarray} \label{mu}
\mu_{+}^{(\sigma)}=-\mu_{-}^{(\sigma)}=-2^{1/8}, \quad \mu_{+}^{(\epsilon)}=\mu_{-}^{(\epsilon)}=-\mu_f^{(\epsilon)}=4 \, .
\end{eqnarray}
The exponent $2-x_{\phi}$ of $y$ in the expansion arises from the scaling dimension $x_T=2$ of $T$. 

For $(\phi, h)= (\sigma , f)$, the leading boundary operator in the expansion (\ref{BOE}) cannot be the stress tensor, as follows from a symmetry argument \cite{symarg}, but has scaling dimension  $1\over 2$, implying the power $y^{1/2 - x_{\sigma}}$.

Although not a primary operator, the expansion (\ref{BOE}) also holds for $\phi(x,y)=T(z)$, with $\langle T\rangle_h=0$, $x_T=2$, and $\mu_h^{(T)}=1$. Due to the analyticity properties of $T(z)$, its expansion contains the powers $y^0$, $y^1$, $y^2$, etc. In averages $\langle T(z)\phi_1\phi_2\dots\rangle_{ab...}$ of $T(z)$ with primary operators, the terms in the expansion can be derived explicitly from the conformal Ward identity, e.g. Eq.~(\ref{GGWI}).
{The boundary-operator expansion (\ref{BOE}) not only applies to the two-dimensional Ising model, but appears to hold quite generally in semi-infinite critical systems, except in the case of a free boundary with $\phi$ equal to the order parameter. This was assumed in Ref.~\cite{Cardydistantwall}, in a study of critical behavior in the parallel-plate geometry.  The asymptotic behavior (\ref{BOE}) has been confirmed in spatial dimension  $d=4-\epsilon$ for the $n$-vector model with $f$ boundary \cite{EEKD,McAvOs,DDE} and for the Ising model with $h=+$ boundary \cite{EEStap}. For $d>2$ , $T(x)$ is replaced by the perpendicular component $T_{yy}$ of the Cartesian stress tensor at the boundary. The expansion is also consistent with a general argument \cite{TWBHWD} that the leading boundary operator for the Ising model in $d$ spatial dimensions with $h=+$ and $\phi=\sigma$ or $\epsilon$ has scaling dimension $d$. Finally, the expansion agrees with the exact results of Ref.~\cite{Cardyscp} for 
$\langle\epsilon_1\epsilon_2\rangle_f=\langle\epsilon_1\epsilon_2\rangle_+$ and of Ref.~\cite{TWBX} for $\langle\sigma\rangle_{ab}$ and $\langle\epsilon\rangle_{ab}$ in the two-dimensional Ising and $Q$-state Potts models. In the two-dimensional models 
\begin{equation} \label{relationmuhphi}
\mu_h^{(\phi)}=-(4x_{\phi}/\hat{c})\,y^{x_\phi}\langle\phi\rangle_h  
\end{equation}
for primary operators, as shown in footnote  \cite{EEJune7}.}
Here $\hat{c}$ is the central charge in the conformal classification \cite{BPZ,CardyD-L}, which equals 1/2 for the Ising model.

The boundary-operator expansion  (\ref{BOE}), with $\langle\phi\rangle_h$ on the left-hand side evaluated for a uniform boundary $h$, has a local character and also holds for mixed $ab..h..$ boundary conditions if, in the small $y$ limit, $\phi(x,y)$ is positioned closer to an interior point of the segment with boundary condition $h$ than its endpoints. In terms of the position $(x,y)$ of $\phi$ and the endpoints $\zeta_j$, $\zeta_{j+1}$ of the segment, this corresponds to $y\ll |x-\zeta_j|$ and $y\ll |x-\zeta_{j+1}|$. 

For the boundary condition $ab..h..$, averaging expansion (\ref{BOE}) leads to 
\begin{equation} 
\langle\phi(x, y)\rangle_{ab..h..}-\langle \phi \rangle_h \to \mu_h^{(\phi)}  \, y^{2-x_{\phi}}  \,\langle T(x)\rangle_{ab..h..}  \, , \quad  y \to 0 \, . \label{avphiabc}
\end{equation}
We have verified that the exact one-point averages of $\sigma$, $\epsilon$  and $T$ with mixed boundary conditions, 
given in Ref.~\cite{TWBX} and in Secs.~\ref{pfpetc} and \ref{mfp} all have this asymptotic behavior.

Boundary operator expansions also provide information on the asymptotic behavior of correlation functions. Consider, for example, the cumulant of $\phi(x,y)$ and a distant operator $\Phi(X,Y)$. According to expansions (\ref{BOE}) and (\ref{avphiabc}),
\begin{eqnarray} \label{twoaway} 
&& \langle\phi(x,y)\Phi(X,Y)\rangle_{ab..h..}-\langle \phi(x,y) \rangle_{ab..h..} \langle\Phi(X,Y)\rangle_{ab..h..}\nonumber \\
&& \qquad \to \mu_h^{(\phi)}\, y^{2-x_{\phi}}\,\left[ \langle T(x)\Phi(X,Y)\rangle_{ab..h..} - \langle T(x) \rangle_{ab..h..} \langle \Phi(X,Y)\rangle_{ab..h..} \right]\label{avphiPhiabc}
\end{eqnarray}
for $y$ much smaller than $|x-\zeta_j|$, $|x-\zeta_{j+1}|$, and $\left[(x-X)^2 + (y-Y)^2\right]^{1/2}$.  The right-hand side of Eq.~(\ref{avphiPhiabc}) can be expressed in terms of $\langle \Phi(X,Y)\rangle_{ab..h..}$ and its derivatives using the conformal Ward identity (\ref{GGWI}).
The asymptotic form (\ref{avphiPhiabc})  is consistent with all the exact expressions for the two-point functions $\langle\sigma_1\sigma_2\rangle$, $\langle\epsilon_1\epsilon_2\rangle$, and $\langle\sigma_1\epsilon_2\rangle$ with mixed boundary conditions given in Ref. \cite{TWBX} and in this paper. For $\langle\sigma_1\sigma_2\rangle_{+-}$ this is shown in some detail in Appendix \ref{appendixcheckaway}.

\subsection{Boundary-operator expansion at a switching point}
Now we turn to operator expansions in the contrasting case in which $\phi(x,y)$ is positioned much closer to one of the switching points, say $\zeta_1$, than to the other switching points $\{ \zeta \} \equiv\zeta_2, \zeta_3,... $ and, when considering multipoint averages, to other operators  $\Phi_1(X_1,Y_1)$, $\Phi_2(X_2,Y_2)$, ... In terms of the complex coordinates $z=x+iy$ and $Z=X+iY$, this corresponds to $|z-\zeta_1| \ll |z-\zeta_2|,\,...\, , |z-Z_1|,\,...$ 
Below, in discussing the order of terms in expansions, we use the notation $l$ and $L$ for small and large lengths, such as $z-\zeta_1$ and $z-\zeta_2$, respectively. 

In leading order the expansion in terms of boundary-operators at the switching point $\zeta_1$ has the form
\begin{eqnarray} \label{MBOE}
\phi(x,y)-\left\langle \phi(x,y) \right\rangle _{ab}^{(\zeta_1)} \to 
F_{ab}^{(\phi)}(x-\zeta_1 ,y) \, \Upsilon({\zeta_1})\,. 
\end{eqnarray}
Here $\phi$ can be either $\sigma, \epsilon$, or $T$, and $a$ and $b$ are the boundary conditions of the segments that extend from $\zeta_1$ to the left and right, respectively. On the right-hand side of Eq.~(\ref{MBOE}) only the contribution of the boundary-operator $\Upsilon(\zeta_1)$ of lowest scaling dimension is shown. Like the factor $\mu_h^{(\phi)}  \, y^{2-x_{\phi}}$ in Eq. (\ref{BOE}), $F_{ab}^{(\phi)}$ in (\ref{MBOE}) only depends on local properties. It depends on the boundary conditions $a, b$ of the two segments with switching point $\zeta_1$ but is independent of any other segments and switching points. According to Eq.~(\ref{MBOE}), $\langle \Upsilon (\zeta_1) \rangle_{ab}^{(\zeta_1)} =0$  if the entire boundary consists of one $a$ segment and one $b$ segment.

As shown in Appendix~\ref{appendixderivation}, for all pairs of universality classes $ab$ the scaling dimension of $\Upsilon$ equals 1, not just for the Ising model, but for other two-dimensional critical systems as well. Thus, the scaling dimension of $F_{ab}^{(\phi)}$ is $x_{\phi} - 1$. The analyticity properties and scaling dimension  $x_T = 2$ of the stress tensor $T(z)$ imply that $F_{ab}^{(T)}$ is proportional to $(z-\zeta_1)^{-1}$, and we normalize $\Upsilon({\zeta_1})$ so that 
\begin{eqnarray} \label{FabT}
F_{ab}^{(T)}(x-\zeta_1,y) = {1 \over z-\zeta_1}\ .\label{FabT}
\end{eqnarray}
In Appendix~\ref{appendixderivation} we show that \begin{equation}
F_{ab}^{(\phi)}(x-\zeta_1,y)=(2t_{ab})^{-1}\,|z-\zeta_1|^2\,\partial_{\zeta_1}\langle\phi\rangle_{ab}\label{generalresult2text}
\end{equation}
for primary operators. Another derivation of this result, based on the conformal Ward identity, is discussed below Eq.~(\ref{Kexpand2}).
We emphasize that expressions (\ref{FabT}) and (\ref{generalresult2text}) are not restricted to the Ising model, but are expected to also hold for other two-dimensional critical systems.

According to Eq. (\ref{MBOE}), the change in $\langle\phi(x,y)\rangle$ near the switching point $\zeta_1$  induced by distant switching points $\{\zeta\}=\zeta_2,\,\zeta_3,...$ has the form 
\begin{eqnarray} \label{MBOEone}
\left\langle \phi(x,y) \right\rangle_{ab\{ c \}}^{(\zeta_1,\{ \zeta \})}-\left\langle \phi(x,y) \right\rangle _{ab}^{(\zeta_1)}  \to 
F_{ab}^{(\phi)}(x-\zeta_1 , y) \, \left\langle\Upsilon({\zeta_1})\right\rangle_{ab\{ c \}}^{(\zeta_1,\{ \zeta \})} \, .
\end{eqnarray}
This complements the change (\ref{avphiabc}) in $\langle\phi(x,y)\rangle$ near {\em interior} points of a boundary segment due to distant switching points. 
In terms of the small and large lengths $l$ and $L$, the leading contribution $\propto l^{- x_{\phi}}$ of the first term on the left-hand side of Eq.~(\ref{MBOEone}) is cancelled by the second term on the left, and the right-hand side of (\ref{MBOEone}), $\propto (l/L) \times l^{- x_{\phi}}$, represents the next-to-leading contribution. On the right-hand side of Eq. (\ref{MBOEone}) the dependence on the distant switching points $\{ \zeta \}$ and the universality classes $\{ c \}$ of the corresponding segments is entirely contained  in the second factor $\left\langle\Upsilon({\zeta_1})\right\rangle _{ab\{ c \}}$, which is independent of  $\phi$. The  dependence on $\phi$ comes from the first factor $F_{ab}^{(\phi)}$, shown in Eqs.~(\ref{FabT}) and (\ref{generalresult2text}), which, as already mentioned, is independent of the distant switching points and their universality classes.

Explicit expressions for $\langle\Upsilon({\zeta_1})\rangle_{ab\{ c \}}$ follow readily from Eqs.~(\ref{FabT}) and (\ref{MBOEone}), with $\phi =T$, which imply
\begin{equation} 
\left\langle T(z)\right\rangle_{ab\{ c \}}^{(\zeta_1,\{ \zeta \})}-\left\langle T(z) \right\rangle _{ab}^{(\zeta_1)}  \to 
{1 \over z-\zeta_1} \, \left\langle\Upsilon({\zeta_1})\right\rangle_{ab\{ c \}}^{(\zeta_1,\{ \zeta \})} \, . \label{MBOET}
\end{equation}
Inserting the stress tensors (\ref{Tab}) and (\ref{Tabc}) for $ab$ and $abc$ boundaries on the left-hand side of (\ref{MBOET}) leads to
\begin{equation} 
\langle \Upsilon ({\zeta_1}) \rangle_{abc}^{(\zeta_1, \zeta_2)} = {t_{ab}+t_{bc}-t_{ac}\over \zeta_2 -\zeta_1} \, ,\label{avupsaba}
\end{equation}
which, like  Eq.~(\ref{Tabc}), holds for $c\neq a$ and $c= a$, with $t_{aa}=0$ in the latter case.  Similarly, from the stress tensors (\ref{T+f+etc}) and (\ref{T+-+etc}) for the Ising model with alternating $+f+f+\dots$ and $+-+-+\dots$ boundary conditions, we obtain
\begin{equation}
\begin{array}{l}
\langle\Upsilon({\zeta_1})\rangle^{(\zeta_1,\dots,\zeta_m)}_{+f+f+...}=\partial_{\zeta_1} {\rm ln} \, G_\sigma^{(m,1)}(\zeta_1,\dots,\zeta_m)\,,\\[1mm]
\langle\Upsilon({\zeta_1})\rangle^{(\zeta_1,\dots,\zeta_m)}_{+-+-+...}=\partial_{\zeta_1} {\rm ln} \, G_\epsilon^{(m)}(\zeta_1,\dots,\zeta_m)\,.
\end{array}\label{avups+f+f+}
\end{equation}

In Appendix~\ref{relationUpsfreeenergy} we show that the quantity $\langle\Upsilon(\zeta_j)\rangle_{abc...}$ has a direct physical interpretation. It can be expressed as a free-energy derivative and represents a fluctuation-induced or Casimir force on switching point $\zeta_j$. In Appendix~\ref{appendixUpsUps} we show that multipoint averages of the boundary operator $\Upsilon$, such as
$\langle\Upsilon(\zeta_1)\Upsilon(\zeta_2)\rangle_{abc...\,}$, are also determined by the operator expansion at a switching point.

For $(\phi,h) \neq (\sigma,f)$, the asymptotic form of $F_{ab}^{(\phi)}$ near an interior point $z=x$ of the $a$ or $b$ interval, i.e., for $y\to 0$, $x\neq \zeta_1$, follows from Eq.~(\ref{MBOEone}), on using Eq. (\ref{BOE}) to express both terms on the left-hand side in terms of the stress tensor. This leads to
\begin{eqnarray}\label{MBOE1} 
\mu_h^{(\phi)}\, y^{2-x_{\phi}} \, \Big[\big\langle T(x) \big\rangle_{ab\{ c \}}^{(\zeta_1,\{ \zeta \})}-\big\langle T(x) \big\rangle _{ab}^{(\zeta_1)} ]\to 
F_{ab}^{(\phi)}(x-\zeta_1 , y) \, \big\langle\Upsilon({\zeta_1})\big\rangle_{ab\{ c \}}^{(\zeta_1,\{ \zeta \})} \, . \end{eqnarray}
Making the substitution (\ref{MBOET}), with $z=x$, in Eq.~(\ref{MBOE1}), we obtain
\begin{eqnarray} \label{Fhom}
F_{ab}^{(\phi)}(x-\zeta_1, y) \to  \mu_h^{(\phi)} \,{y^{2-x_{\phi}}\over x-\zeta_1}\, ; \quad y\to 0\,,\; x \neq \zeta_1\,.\label{Fabasymp}
\end{eqnarray}
This result holds for $\phi=\sigma$, $\epsilon$ and $T$, with $h=a$ for $x < \zeta_1$ and $h=b$ for 
$x > \zeta_1$, provided $(\phi,h) \neq (\sigma,f)$.  The amplitudes $\mu_h^{(\phi)}$ are given in and just below Eq. (\ref{mu}). The functions $F_{ab}^{(\phi)}$ for $\phi=\sigma$ and $\epsilon$ are determined explicitly for the Ising model in Subsec.~\ref{abboundaries}  (see Eqs.~(\ref{dsigandepsabdzeta1}) and (\ref{Fabphi2}) and do indeed have the asymptotic behavior (\ref{Fabasymp}), as does $F_{ab}^{(T)}$ in Eq. (\ref{FabT}), with $x_T=2$ and $\mu_h^{(T)}=1$.

The operator expansion (\ref{MBOE}) also yields asymptotic information on averages $\langle \phi\,  \Phi_1 \Phi_2 ... \rangle_{ab\{ c \}}^{(\zeta_1,\{ \zeta \})}$ of products of an operator $\phi$ positioned close to the switching point $\zeta_1$ and distant operators $\Phi_1,\Phi_2,..$. We study this in detail for two-point averages, where 
Eq. (\ref{MBOE}) leads to
\begin{eqnarray} \label{MBOEtwo}
&&\langle \phi(x,y) \Phi(X,Y) \big\rangle _{ab\{ c \}}^{(\zeta_1,\{ \zeta \})}-\big\langle \phi(x,y) \rangle_{ab}^{(\zeta_1)} \langle \Phi(X,Y) \rangle_{ab\{ c \}}^{(\zeta_1,\{ \zeta \})} \nonumber \\
&& \to F_{ab}^{(\phi)}(x-\zeta_1 , y)\,\left\langle \Upsilon(\zeta_1) \Phi(X,Y) \right\rangle_{ab\{ c \}}^{(\zeta_1,\{ \zeta \})} \, .
\end{eqnarray}
In our further analysis we decompose the average on the right-hand side of Eq. (\ref{MBOEtwo}) according to
\begin{eqnarray} \label{UpsD}
\langle \Upsilon(\zeta_1) \Phi(X,Y) \rangle_{ab\{c\}}^{(\zeta_1,\{ \zeta \})} = \bigl[ \langle \Upsilon(\zeta_1) \rangle_{ab\{c\}}^{(\zeta_1,\{ \zeta \})} + \partial_{\zeta_1} \bigr] \langle \Phi(X,Y) \rangle_{ab\{c\}}^{(\zeta_1,\{ \zeta \})} 
\end{eqnarray}
where the derivative $\partial_{\zeta_1}$ is at fixed $X,Y,\{ \zeta \}$. This  relation is consistent with the exact results for one and two-point averages with mixed boundary conditions in Refs.~\cite{TWBX,TWBG2} and in Secs.~\ref{pfpetc} and \ref{mfp} of this paper. In addition, the scaling dimension 1 of $\Upsilon$ allows for the first derivative of a length, and, due to locality, only $\zeta_1$ qualifies. Finally, the term with derivative $\partial_{\zeta_1}$ and with a prefactor of 1 in Eq.~({\ref{UpsD}) follows from a  conformal Ward identity for $\Phi$, as we discuss below Eq. (\ref{GGWI}). 

For convenience we often omit the superscripts $(\zeta_1)$ and $(\zeta_1,\{\zeta\})$ below.
Substituting Eq. ~(\ref{UpsD}),  
into Eq. (\ref{MBOEtwo}) leads to
\begin{eqnarray} \label{explicitnextlead} 
\langle \phi \Phi\rangle_{ab\{ c \}} -\langle \phi  \rangle_{ab}  \langle \Phi  \rangle_{ab\{ c \}}
\to F_{ab}^{(\phi)}\times \left[\langle \Upsilon\rangle_{ab\{ c \}}+\partial_{\zeta_1} \right]\langle \Phi \big\rangle_{ab\{ c \}}\,.
\end{eqnarray}
In analogy with Eq.~(\ref{MBOEone}), the leading contribution, $\propto l^{- x_\phi} L^{-x_\Phi}$, of the first term on the left-hand side of Eq.~{\ref{explicitnextlead}) is cancelled by the second term on the left, and the right-hand side, $\propto (l/L) \times l^{- x_\phi}L^{-x_\Phi}$, represents the next-to-leading contribution. 
Combining Eqs.~(\ref{MBOEone}) and (\ref{explicitnextlead}), we obtain

\begin{eqnarray} \label{kumu}
\langle\phi(x,y)\Phi(X,Y)\rangle_{ab\{ c \}}^{\rm cum}& \equiv & \langle \phi \Phi\rangle _{ab\{ c \}} - \big\langle \phi \big\rangle_{ab\{ c \}} \langle \Phi \rangle_{ab\{ c \}}\nonumber\\ &\to&  
F_{ab}^{(\phi)}(x-\zeta_1,y)\,\partial_{\zeta_1}\langle\Phi(X,Y)\rangle_{ab\{ c \}}
\end{eqnarray}
for the asymptotic form of the cumulant of $\phi$ and $\Phi$. 
On substituting Eqs.~(\ref{generalresult2text}) and (\ref{FabT}), Eq.~(\ref{kumu}) takes the form 
\begin{equation}
\langle\phi\Phi\rangle_{ab\{ c \}}^{\rm cum}  \to\left\{ \begin{array}{l}  
(2t_{ab})^{-1}\,|z-\zeta_1|^2\;\partial_{\zeta_1}\langle\phi\rangle_{ab}\;\partial_{\zeta_1}\langle\Phi\rangle_{ab\{ c \}}\,,\\
(z-\zeta_1)^{-1}\,\partial_{\zeta_1}\langle\Phi\rangle_{ab\{ c \}}\,,\end{array}\right.\;
\begin{array}{l} \phi=\sigma{\rm\;or\;\epsilon}\,, \\ \phi=T\,,\end{array}\label{kumunew}
\end{equation}
in terms of derivatives of one-point averages. As a consequence, ratios $\langle\phi\Phi_1 \rangle_{ab\{ c \}}^{\rm cum} / \langle\phi\Phi_2 \rangle_{ab\{ c \}}^{\rm cum}$ of cumulants with different $\Phi$'s but the same $\phi$ are independent of $\phi$, and vice versa.

As a check on Eqs.~(\ref{kumu}) and (\ref{kumunew}), we recall the conformal Ward identity \cite{Cardyscp,TWBX}
\begin{eqnarray} \label{GGWI}
&&\left\langle T(z) \Phi(X,Y)\right\rangle_{ab\{c\}}^{(\zeta_1, \{\zeta\})} -\left\langle T(z) \right\rangle_{ab\{c\}}^{(\zeta_1,\{\zeta\})}  \langle \Phi(X,Y)\rangle_{ab\{c\}}^{(\zeta_1,\{\zeta\})}
= \left[ {\Delta_\Phi \over (z-Z)^2}\right. \nonumber\\ &&\quad \left.+ {1 \over z -Z}\,{\partial \over \partial Z} + {\Delta_\Phi\over (z-\bar{Z})^2} + {1 \over z-\bar{Z}} \,{\partial \over \partial \bar{Z}} +  \sum _{j}{1 \over z-\zeta_j} {\partial \over \partial \zeta_j} \right] \langle \Phi(X,Y) \rangle_{ab\{c\}}^{(\zeta_1,\{\zeta\})}\,,\quad 
\end{eqnarray}
where $\Phi(X,Y)$ is a primary operator. In the limit in which $z$ is much closer to $\zeta_1$ than to any other of the switching points and to $Z$,
all the terms on the right-hand side of Eq.~(\ref{GGWI}) are of order $L^{-2-x_\Phi}$ except the term $(z -\zeta_1)^{-1}\partial_{\zeta_1}  \langle \Phi(X,Y) \rangle$, which is of order $(L/l) L^{-2-x_\Phi}$, and thus the leading contribution. Making use of Eq.~(\ref{FabT}), we see that the leading contribution  is the same as the asymptotic forms of the cumulant $\langle\phi\Phi\rangle_{ab\{c\}}^{\rm cum}$ in Eqs.~(\ref{kumu}) and (\ref{kumunew}) for $\phi=T$. This validates the prediction of the operator expansion for $\phi=T$ and for $\Phi$ equal to a primary operator, such as $\sigma$ or $\epsilon$ in the Ising model.
 
In the remainder of this section we specialize to $ab$ and $abc$ boundaries. In Subsec.~\ref{abboundaries} the consistency of the asymptotic forms ~(\ref{kumu}) and (\ref{kumunew}) with Ward identities and with exact results for $\langle\phi\Phi\rangle_{ab}^{\rm cum}$ in the Ising model is checked.  Similar consistency checks are carried out for $abc$ boundaries in Subsec.~\ref{abcboundaries}.

\subsection{$ab$ boundaries}\label{abboundaries}

In this subsection we first confirm, with the help of Ward identities, that the asymptotic form of the two-point cumulant in Eqs.~(\ref{kumu}) and (\ref{kumunew}) holds if either $\phi$ or $\Phi$ or both equal $T$. Then we derive the functions $F_{ab}^{(\phi)}$, $\partial_{\zeta_1}\langle\phi\rangle_{ab}\,$ and $\partial_{\zeta_1}\langle\Phi\rangle_{ab}\,$ on the right-hand sides of Eqs.~(\ref{kumu}) and (\ref{kumunew}) explicitly for the Ising model and confirm the consistency of the predicted asymptotic behavior with exact results for the two-point averages.

\subsubsection{Confirmation of the asymptotic form (\ref{kumu}) for $\phi$ or $\Phi$ or both equal to $T$}

Beginning with the Ward identity (\ref{GGWI}), we already showed that Eq.~(\ref{kumu}) holds for $\phi=T$ and $\Phi$ equal to a primary operator. It also holds for  $\phi=\Phi=T$, since substituting  Eqs.~(\ref{FabT}) and its derivative
\begin{equation}
\partial_{\zeta_1}\langle T(Z)\rangle_{ab}={2t_{ab}\over (Z-\zeta_1)^3}  \label{dTabdzeta1}
\end{equation} 
in Eq.~(\ref{kumu}) leads to
\begin{equation}
\langle T(z)T(Z)\rangle_{ab}^{\rm cum}\to{2t_{ab}\over(z-\zeta_1)(Z-\zeta_1)^3}\,,\label{TTcumasymp}
\end{equation}
which agrees with the exact result for $\langle T(z)T(Z)\rangle_{ab}^{\rm cum}$ discussed in Appendix~\ref{appendixTT} and shown in Eq.~(\ref{TTcumab}), in the limit that $z$ is much closer to $\zeta_1$ than to $Z$.

We now consider the cumulant $\langle\phi T\rangle_{ab}^{\rm cum}$ for $\phi$ equal to a primary operator and show its consistency  with Eq.~{(\ref{kumu}). The starting point is the conformal Ward identity
\begin{eqnarray} \label{GGWII}
&&\left\langle T(Z) \phi(x,y)\right\rangle_{ab}^{(\zeta_1)}-\left\langle T(Z) \right\rangle_{ab}^{(\zeta_1)}  \left\langle \phi(x,y)\right\rangle_{ab}^{(\zeta_1)}
= \left[ {\Delta_\phi \over (Z-z)^2} \right.\nonumber\\ &&\qquad \left.+ {1 \over Z-z}\,{\partial \over \partial z} + {\Delta_\phi\over (Z-\bar{z})^2} + {1 \over Z-\bar{z}} \,{\partial \over \partial \bar{z}} +  {1\over Z-\zeta_1}\,{\partial \over \partial \zeta_1} \right] \left\langle \phi(x,y) \right\rangle_{ab}^{(\zeta_1)}\,,\quad 
\end{eqnarray}
which is the same as Eq.~(\ref{GGWI}), except that $\phi$ and $\Phi$, $z$ and $Z$, and $\bar{z}$ and  $\bar{Z}$ have been exchanged, and we specialize to an $ab$ boundary with a single switching point $\zeta_1$.
Noting that the left-hand side of Eq.~(\ref{GGWII}) is $\langle\phi T\rangle_{ab}^{\rm cum}$ and expanding the $z$ and $\bar z$ dependence of the square bracket in a Taylor series about $z=\bar z=\zeta_1$ leads to 
\begin{eqnarray} \label{KOT}
&&\langle\phi T\rangle_{ab}^{\rm cum} \to\left\{(Z-\zeta_1)^{-1} \left(\partial_{z} + \partial_{\bar{z}}+\partial_{\zeta_1}\right)
+ (Z-\zeta_1)^{-2} \left(x_\phi + \delta z\,\partial_{z} + \delta\bar{z}\, \partial_{\bar{z}} \right)\right.   \nonumber \\
&& \quad \left.+ (Z-\zeta_1)^{-3} \left[x_\phi(\delta z+ \delta\bar{z})+ (\delta z)^2\, \partial_{z} + (\delta\bar{z})^2\, \partial_{\bar{z}} \right]+\dots\right\} 
\langle \phi(x,y) \rangle_{ab}^{(\zeta_1)} \, .\label{Kexpand1}
\end{eqnarray}
where $\delta z\equiv z-\zeta_1$ and $x_\phi=2\Delta_\phi$. The terms $\propto (Z-\zeta_1)^{-1}$ and  $\propto (Z-\zeta_1)^{-2}$ vanish due to the translational  and dilatational invariance \cite{dilatation}, respectively,  of $\langle \phi(x,y) \rangle_{ab}^{(\zeta_1)} $. Using dilatation invariance to replace $x_\phi$ by $-(\delta z\partial_z+\delta\bar z\partial_{\bar z})$ in the term $\propto (Z-\zeta_1)^{-3}$, we obtain
\begin{eqnarray} \label{KOT}
\langle\phi T\rangle_{ab}^{\rm cum} \to  -(Z&-&\zeta_1)^{-3}\, \delta z \delta\bar{z}\,(\partial_{z} + \partial_{\bar{z}})\,
\langle \phi(x,y) \rangle_{ab}^{(\zeta_1)} \nonumber \\
&&\quad= (Z-\zeta_1)^{-3}\, \bigl\vert z-\zeta_1\bigr\vert ^2\,\partial_{\zeta_1} \langle \phi(x,y) \rangle_{ab}^{(\zeta_1)}\nonumber\\
&&\quad=(2t_{ab})^{-1}\,|z-\zeta_1|^2\,\partial_{\zeta_1}\langle\phi(x,y)\rangle_{ab}\,\partial_{\zeta_1}\langle T(Z)\rangle_{ab}\,.\label{Kexpand2}
\end{eqnarray}
to leading, non-vanishing order.
Here, in going from the first line to the second, we have used translational invariance to replace $\partial_{z} + \partial_{\bar{z}}$ by $-\partial_{\zeta_1}$ and the definition of $\delta z$. Then the second line was rewritten, using Eqs.~(\ref{dTabdzeta1}), to obtain the third line. 

The third line line of Eq.~(\ref{Kexpand2}) is in complete agreement with the asymptotic form (\ref{kumunew}) of $\langle\phi\Phi\rangle_{ab}$ for $\Phi=T$ predicted by the boundary-operator expansion. For consistency with the alternate asymptotic form (\ref{kumu}),  $F_{ab}^{(\phi)}$ and $\partial_{\zeta_1}\langle\phi\rangle_{ab}$} must satisfy Eq.~(\ref{generalresult2text}). This provides an alternate derivation of that relation. 

The results of the paragraph containing  Eq.~(\ref{TTcumasymp}) and Eq.~(\ref{Kexpand2}) confirm the prediction (\ref{kumu}) of the boundary-operator expansion at $\zeta_1$ for $\phi$ or $\Phi$ or both equal to $T$.

\subsubsection{Explicit expressions for $F_{ab}^{(\phi)}$, $\partial_{\zeta_1}\langle\phi\rangle_{ab}$,  and $\partial_{\zeta_1}\langle\Phi\rangle_{ab}$ in the Ising model}\label{explicitab}

Our notation $ab$ for the boundary, i.e., $a$ for $x<\zeta_1$ and $b$ for $x>\zeta_1$, corresponds to $ba$ in the notion of Ref.~\cite{TWBX}. Expressed in our notation, the Ising one-point averages in Eq.~(4.1) of  Ref.~\cite{TWBX} read
\begin{equation} 
\begin{array}{l}\langle\sigma\rangle_{+-}=-\langle\sigma\rangle_{-+}=-\langle\sigma\rangle_{_{+}}^{(y)}\,\cos{\vartheta}\,,\\
\langle\epsilon\rangle_{+-}=\langle\epsilon\rangle_{-+}=\langle\epsilon\rangle_{_{+}}^{(y)}\,(1-4\sin^2\vartheta),\\  
\langle\sigma\rangle_{+f}=\langle\sigma\rangle_{_{+}}^{(y)}\,\left(\sin{\vartheta\over 2}\right)^{1/2}\,,\\
\langle\sigma\rangle_{f+}=\langle\sigma\rangle_{_{+}}^{(y)}\,\left(\cos{\vartheta\over 2}\right)^{1/2}\,,\\
\langle\epsilon\rangle_{+f}=-\langle\epsilon\rangle_{f+}=-\langle\epsilon\rangle_{_{+}}^{(y)}\,\cos\vartheta\,,
\end{array}\label{sigandepsab}
\end{equation}
where $\langle\sigma\rangle_{+}^{(y)}=(2/y)^{1/8}$
and  $\langle\epsilon\rangle_{+}^{(y)}=-(2y)^{-1}$ are the averages for a uniform, spin-up boundary given in Eqs.~(\ref{sigfixed}) and (\ref{epsfixedfree}). Here and below, $(r,\vartheta)$ and $(R,\Theta)$ are polar coordinates defined by 
\begin{eqnarray} 
(x-\zeta_1 , y) = r(\cos \vartheta, \sin \vartheta) \, , \quad
(X-\zeta_1 , Y) = R(\cos \Theta, \sin \Theta)\,. \label{rthetaRTheta}
\end{eqnarray}

Differentiating Eq.~(\ref{sigandepsab}), using $\partial_{\zeta_1}\vartheta=\partial_{\zeta_1}\arctan\left[y/(x-\zeta_1)\right]=r^{-1}\sin\vartheta$, leads to
\begin{equation}\label{dsigandepsabdzeta1}
\begin{array}{l}
\partial_{\zeta_1}\langle\sigma\rangle_{+-}=-\partial_{\zeta_1}\langle\sigma\rangle_{-+}=\langle\sigma\rangle_{+}^{(y)}\,r^{-1}\sin^2\vartheta\,,\\
\partial_{\zeta_1}\langle\epsilon\rangle_{+-}=\partial_{\zeta_1}\langle\epsilon\rangle_{-+}=-8\langle\epsilon\rangle_{+}^{(y)}\, r^{-1}\sin^2\vartheta\cos\vartheta\,,\\
\partial_{\zeta_1}\langle\sigma\rangle_{+f}=
\textstyle{1\over 2}\langle\sigma\rangle_{+}^{(y)}\,r^{-1}\left(\sin{\vartheta\over 2}\right)^{1/2}(\cos{\vartheta\over2})^2\,,\\ 
\partial_{\zeta_1}\langle\sigma\rangle_{f+}=\textstyle -{1\over 2}\langle\sigma\rangle_{+}^{(y)}\,r^{-1}\left(\cos{\vartheta\over 2}\right)^{1/2}(\sin{\vartheta\over2})^2\,,\\
\partial_{\zeta_1}\langle\epsilon\rangle_{+f}=-\partial_{\zeta_1}\langle\epsilon\rangle_{f+}=\langle\epsilon\rangle_{+}^{(y)}\, r^{-1}\,\sin^2\vartheta\,.
\end{array}
\end{equation}
	
The functions $F_{ab}^{(\phi)}$ are easily obtained from these results using Eq.~(\ref{generalresult2text}) in the form
\begin{equation}
F_{ab}^{(\phi)}=(2t_{ab})^{-1} r^2\,\partial_{\zeta_1}\langle\phi\rangle_{ab}\label{Fabphi2}
\end{equation}
and the values $t_{+-}={1\over 2}$ and $t_{+f}={1\over 16}$, given below Eq.~(\ref{Tab}). Thus, for example, $F_{+f}^{(\epsilon)}=8\langle\epsilon\rangle_{+}^{(y)}\, r\, \sin^2\vartheta\,.$  It is simple to check that the expressions for $F_{ab}^{(\phi)}$ are indeed consistent with the asymptotic form (\ref{Fabasymp}) for $y\to 0$, $x\neq\zeta_1$. 

 The quantities $\partial_{\zeta_1}\langle\Phi\rangle_{ab}$ with $\Phi=\sigma$ or $\epsilon$ are the same as in Eq.~(\ref{dsigandepsabdzeta1}), except that $r$, $\vartheta$, and $y$ are replaced $R$, $\Theta$, and $Y$.

Using the explicit expressions $F_{ab}^{(\phi)}$, $\partial_{\zeta_1}\langle\phi\rangle_{ab}$, and $\partial_{\zeta_1}\langle\Phi\rangle_{ab}\,$, we have compared the asymptotic form (\ref{kumu}) or (\ref{kumunew}) of $\langle\phi\Phi\rangle_{ab}$, predicted by the boundary-operator expansion with the asymptotic behavior of the exact two-point functions $\langle\sigma_1\sigma_2\rangle_{ab}$, 
$\langle\sigma_1\epsilon_2\rangle_{ab}$, and $\langle\epsilon_1\epsilon_2\rangle_{ab}$ for $ab=+-$ and $+f$ in Eq.~(4.3) of Ref.~\cite{TWBX} and found complete agreement. In Appendix~\ref{appendixcheckat}, the consistency check is illustrated for $\phi=\Phi=\sigma$ and $ab=+-$ in some detail.

\subsection{$abc$ boundaries}\label{abcboundaries} 
For $abc$ boundaries the asymptotic behavior of one and two-point averages near the switching point $\zeta_1$ is specified by Eqs.~(\ref{MBOEone}) and (\ref{kumu}) or (\ref{kumunew}). In this subsection we first confirm, with the help of Ward identities, that the asymptotic form (\ref{kumu}) holds if either $\phi$ or $\Phi$ or both equal $T$. Then we determine the various functions on the right hand sides of Eqs.~(\ref{MBOEone}) and (\ref{kumu}) explicitly, for the Ising model with $abc$ boundaries. Finally, we confirm the consistency of the predicted asymptotic behavior with exact results for the Ising model.

\subsubsection{Confirmation of the asymptotic form (\ref{kumu}) for $\phi$ or $\Phi$ or both equal to $T$}

We begin by differentiating the stress tensor for $abc$ boundaries (\ref{Tabc}) with respect to $\zeta_1$. This leads to
\begin{equation}
\partial_{\zeta_1}\langle T(Z)\rangle_{abc}={2t_{ab}\over (Z-\zeta_1)^3}+{t_{ac}-t_{ab}-t_{bc}\over (Z-\zeta_1)^2(Z-\zeta_2)}\,,\label{dTabcdzeta1}
\end{equation}
a result we will need below. Like Eq.~(\ref{Tabc}), it holds for $c\neq a$ and $c=a$, with $t_{aa}=0$ in the latter case.

The general argument presented below the Ward identity (\ref{GGWI}), that the cumulant expression (\ref{kumu}) holds for $\phi=T$ and $\Phi$ equal to primary operators, includes the case of $abc$ boundaries. Equation~({\ref{kumu}) also holds when
both $\phi$ and $\Phi$ equal $T$, since $\langle T(z)T(Z)\rangle_{abc}^{\rm cum}\to F_{ab}^{(T)} \times\partial_{\zeta_1}\langle T\rangle_{abc}$, with the right-hand side given by Eqs.~(\ref{FabT}) and (\ref{dTabcdzeta1}), agrees with the exact result for $\langle T(z)T(Z)\rangle_{abc}^{\rm cum}$ discussed in Appendix~\ref{appendixTT} and shown in Eq.~(\ref{TTcumabc}),
in the limit that $z$ is much closer to $\zeta_1$ than to $\zeta_2$ and to $Z$.

Next we confirm Eq.~(\ref{kumu}) for $\phi$ equal to a primary operator and $\Phi=T$, modifying Eq.~(\ref{GGWII}) and the steps below it for an $abc$ instead of an $ab$ boundary boundary. The Ward identity is similar to Eq. (\ref{GGWII}), but with an extra term $(Z-\zeta_2)^{-1}\partial_{\zeta_2}$ in the square bracket. In the relations 
$(\partial_z + \partial_{\bar{z}} +\partial_{\zeta_1} +\partial_{\zeta_2})\langle\phi\rangle_{abc}  =0$ and $\left[x_\phi + \delta z\, \partial_z + \delta\bar z\,\partial_{\bar z} + (\zeta_2 -\zeta_1) \partial_{\zeta_2} \right]\langle\phi\rangle_{abc} =0$, corresponding to translational and dilatational invariance \cite{dilatation}, there are also extra terms involving $\partial_{\zeta_2}$. The expansion in Eq.~(\ref{Kexpand2}) is replaced by
\begin{eqnarray} 
&&\langle\phi T\rangle_{abc}^{\rm cum}=\left\{(Z-\zeta_2)^{-1}\partial_{\zeta_2}-(Z-\zeta_1)^{-1}\partial_{\zeta_2}-(Z-\zeta_1)^{-2}(\zeta_2-\zeta_1)\,\partial_{\zeta_2}\right.\nonumber\\
&&\qquad\left. -(Z-\zeta_1)^{-3}\left[\delta z\,\delta\bar z
(\partial_z+\partial_{\bar z})+(\delta z+\delta\bar z)(\zeta_2-\zeta_1)\,\partial_{\zeta_2}\right]+\dots\right\}\langle\phi\rangle_{abc}\,.\label{Kexpand1abc}
 \end{eqnarray}
Substituting $\langle\phi\rangle_{abc}-\langle\phi\rangle_{ab}\to F_{ab}^{(\phi)}\times\langle\Upsilon\rangle_{abc}\,$, which follows from Eq.~(\ref{MBOEone}), and expression (\ref{avupsaba}) for $\langle\Upsilon\rangle_{abc}$, we obtain
\begin{eqnarray} \label{K_ABC_OT'}
\langle\phi T\rangle_{abc}^{\rm cum}&\to& F_{ab}^{(\phi)} \times \left[ \partial_{\zeta_1}\langle T(Z)\rangle_{ab} + {(\zeta_2 -\zeta_1)^2 \over (Z-\zeta_1)^2 (Z-\zeta_2)} \partial_{\zeta_2} \langle \Upsilon(\zeta_1) \rangle _{abc}^{(\zeta_1 , \zeta_2)}  \right]   \nonumber \\
&\to& F_{ab}^{(\phi)} \times \partial_{\zeta_1} \langle T(Z) \rangle _{abc}^{(\zeta_1, \zeta_2)}\,,
\end{eqnarray}
to leading order $(l/L)l^{-x_\phi} L^{-x_T}$.
In going from the first line to the second, we  have used expressions (\ref{avupsaba}), (\ref{dTabdzeta1}), and (\ref{dTabcdzeta1}) for $\langle \Upsilon\rangle _{abc}$, 
$\partial_{\zeta_1}\langle T(Z) \rangle _{ab}$, and $\partial_{\zeta_1}\langle T(Z) \rangle _{abc}$, respectively.

Equation~(\ref{kumu}) with $\Phi=T$ and Eq.~(\ref{K_ABC_OT'}) are clearly consistent. Together with the results discussed below Eq.~(\ref{dTabcdzeta1}), this confirms the asymptotic form (\ref{kumu}) of the two-point cumulant  $\langle\phi\Phi\rangle_{abc}^{\rm cum}$ for either $\phi$ or $\Phi$ or both equal to $T$.

\subsubsection{Explicit expressions for $\partial_{\zeta_1}\langle\Phi\rangle_{abc}$ in the Ising model} 

The explicit form of $\partial_{\zeta_1}\langle T\rangle_{abc}$ is shown in Eq.~(\ref{dTabcdzeta1}). 
Here we consider $\partial_{\zeta_1}\langle\Phi\rangle_{abc}$  for $\Phi=\sigma$ and $\epsilon$ and $abc=+f+$, $+-+$, $-f+$, and  $f+-$, and obtain explicit expressions by differentiating  the  corresponding Ising one-point averages $\langle\sigma\rangle_{abc}$ and $\langle\epsilon\rangle_{abc}$. 
For $+f+$, we begin with the one-point averages $\langle\sigma\rangle_{+f+}$ and $\langle\epsilon\rangle_{+f+}$ in Eqs.~(\ref{sigpfp}) and (\ref{epspfp}), replace $(x_1,y_1)$ by $(X,Y)$ and $\gamma_{1,1}$ by $\Gamma$, where, according to Eq.~(\ref{gammakelldef2}),
\begin{equation}
\textstyle\Gamma={\rm arg}\,{Z-\zeta_2\over Z-\zeta_1}\,,\label{defineGamma}
\end{equation}
and then evaluate the derivative with respect to $\zeta_1$, using 
\begin{equation}
\textstyle\partial_{\zeta_1}\Gamma=-\partial_{\zeta_1}\, \arctan{Y\over X-\zeta_1}=-R^{-1}\sin\Theta\,,\label{gammaderivative}
\end{equation}
which follows from Eqs.~(\ref{rthetaRTheta}) and (\ref{defineGamma}). For $+-+$ boundaries, the calculation is similar, but begins with the one-point averages 
$\langle\sigma\rangle_{+-+}=\langle\sigma\rangle_{_+}^{(Y)}\,\cos\Gamma$
and $\langle\epsilon\rangle_{+-+}=\langle\epsilon\rangle_{_+}^{(Y)}\left(1-4\sin^2\Gamma\right)$, given in \cite{TWBG2} or obtained with the conformal transformation (\ref{abtoaba}) from the results for a $+-$ boundary shown in Eq.~(\ref{sigandepsab}).
For $-f+$ and $f+-$ boundaries the calculations are also similar, but begin with Eqs.~(\ref{sigmfp}), (\ref{epsmfp}), (\ref{sigfpm}), and (\ref{epsfpm}). In this way we obtain


\begin{equation}
\begin{array}{l}
\partial_{\zeta_1}\langle\sigma\rangle_{+f+}={1\over 4}\,\langle\sigma\rangle_{_+}^{(Y)}\,{\sin{\textstyle{\Gamma\over 2}}\over{\sqrt{\cos{\textstyle{\Gamma\over 2}}}}}\,{\sin\Theta\over R}\,,\\[3mm]
\partial_{\zeta_1}\langle\sigma\rangle_{f+f}=-{1\over 4}\,\langle\sigma\rangle_{_+}^{(Y)}\,{\cos{\textstyle{\Gamma\over 2}}\over{\sqrt{\sin{\textstyle{\Gamma\over 2}}}}}\,{\sin\Theta\over R}\,,\\[2mm]
\partial_{\zeta_1}\langle\epsilon\rangle_{+f+}=-\partial_{\zeta_1}\langle\epsilon\rangle_{f+f}=\langle\epsilon\rangle_{_+}^{(Y)}\,\sin\Gamma\,{\sin\Theta\over R}\,,\\[2mm]
\partial_{\zeta_1}\langle\sigma\rangle_{+-+}=\langle\sigma\rangle_{_ +}^{(Y)}\,\sin\Gamma\,{\sin\Theta\over R}\,,\\[2mm]
\partial_{\zeta_1}\langle\epsilon\rangle_{+-+}=8\,\langle\epsilon\rangle_{_+}^{(Y)}\sin\Gamma\cos\Gamma\,{\sin\Theta\over R}\,,\\[2mm]
\partial_{\zeta_1}\langle\sigma\rangle_{-f+}=\langle\sigma\rangle_{_ +}^{(Y)}\,{1\over 4W_1}\left[(1-{4R^2\over\zeta_{21}^2})\sin{\textstyle{\Gamma\over 2}}+{2R\over\zeta_{21}}\sin\Theta\cos{\textstyle{\Gamma\over 2}}\right]\,{\sin\Theta\over R}\,,\\[2mm]
\partial_{\zeta_1}\langle\epsilon\rangle_{-f+}=\langle\epsilon\rangle_{_+}^{(Y)}\,\left[(1-{4R^2\over \zeta_{21}^2})\sin\Gamma+{4R\over\zeta_{21}}\sin\Theta\cos\Gamma\right]\,{\sin\Theta\over R}\,,\\[2mm]
\partial_{\zeta_1}\langle\sigma\rangle_{f+-}=\langle\sigma\rangle_{_+}^{(Y)}{1\over 4W_2}\left[{4R(R-\zeta_{21}\cos^2{\textstyle{\Theta\over 2}})\over\left\vert Z-\zeta_2\right\vert^2}-1\right]{\sin{\textstyle{\Theta\over 2}}\,\sin\Theta\over R}\,,\\[2mm]
\partial_{\zeta_1}\langle\epsilon\rangle_{f+-}=\langle\epsilon\rangle_{_+}^{(Y)}\left[{4R\left(R-\zeta_{21}\cos\Theta\right)\over\left\vert Z-\zeta_2\right\vert^2}-1\right]{\sin^2\Theta\over R}\,.
\end{array}
\label{dPhiabcdzeta1}
\end{equation}
Here  $W_1$ and $W_2$ are the square roots $W_1\equiv\left[\cos{\textstyle{\Gamma\over 2}}-(2Y/\zeta_{21})\sin{\textstyle{\Gamma\over 2}}\right]^{1/2}$ and 
$W_2=\left[\cos{\textstyle{\Theta\over 2}}-\,2\,\zeta_{21}R\left\vert Z-\zeta_2\right\vert^{-2}
\sin{\textstyle{\Theta\over  2}}\sin\Theta\right]^{1/2}$ in Eqs.~(\ref{sigmfp}) and (\ref{sigfpm}), respectively.
The trigonometric functions of $\Gamma$ in Eq.~(\ref{dPhiabcdzeta1}) can be expressed in terms of the Cartesian coordinates $X,Y$ using the relations
\begin{equation}
\cos\Gamma=\displaystyle{(X-\zeta_1)(X-\zeta_2)+Y^2\over R\vert Z-\zeta_2\vert}\,,\quad\sin\Gamma={Y(\zeta_2-\zeta_1)\over R\vert Z-\zeta_2\vert}\,,
\label{trigGamma}
\end{equation}
which follow from Eq.~(\ref{defineGamma}) and correspond to Eqs.~(\ref{cosgammakell}) and (\ref{singammakell}).

Using the explicit expressions for $F_{ab}^{(\phi)}$ and $\partial_{\zeta_1}\langle\Phi\rangle_{abc}$, given in Eqs.~(\ref{dsigandepsabdzeta1}), (\ref{Fabphi2}), and (\ref{dPhiabcdzeta1}), we have confirmed the consistency of the asymptotic behavior of the one and two-point averages, $\langle\phi\rangle_{abc}$ and $\langle\phi\Phi\rangle_{abc}$, shown in Eqs.~(\ref{MBOEone}) and (\ref{kumu}), respectively, with the exact results reported in Secs.~\ref{pfpetc} and \ref{mfp}. In Appendix~\ref{appendixcheckat} the consistency check for $\phi=\Phi=\sigma$ and $abc=+f+$ is carried out in some detail.

\subsection{Distant-wall effects}

At criticality, local behavior throughout the system is affected by the boundaries, even if they are distant. In a classic paper Fisher and de Gennes \cite{FdG} considered a critical fluid confined between infinite parallel plates or walls with separation ${\cal W}$.  Calculating the density profile by minimizing a local free energy functional, they found that the correction to the profile near one wall due to the distant wall varies as ${\cal W}^{-d}$, where $d$ is the spatial dimension

The two-dimensional analog of the fluid between plates is an Ising strip of infinite length and width ${\cal W}$. Exact results for $\langle\sigma\rangle_{a\vert b}$ and
$\langle\epsilon\rangle_{a\vert b}$,  for boundary condition $a$ on one edge and $b$ on the other, obtained by conformally mapping the semi-infinite results (\ref{sigandepsab}) onto the strip geometry, confirm the ${\cal W}^{-2}$ variation of the distant-wall corrections, similar results were obtained for Potts spins, and a general connection in two-dimensional critical systems between the distant-wall corrections to the profiles and the Casimir force between the edges was explained in terms of   conformal invariance \cite{TWBX,Cardydistantwall}.

In these and other studies of distant-wall corrections
(see \cite{TWBX,EEKD,RudnickJasnow,Upton} and references therein), the boundary condition on each wall is assumed to be uniform.
Here we consider distant-wall effects in the critical Ising model defined on an infinitely long  strip with {\it mixed} boundary conditions, thereby demonstrating the versatility of the boundary-operator approach. The lower boundary of the strip is the $x$ axis, and the upper boundary is parallel to the $x$ axis and a distance ${\cal W}$ above it. Imposing $ab\vert c$ boundary conditions, consisting of $ab$ boundary conditions with switching point $\zeta_1$ on the lower boundary and a uniform boundary condition $c$ on the upper boundary, we analyze the effect of the distant upper boundary on the profile $\langle\phi(x,y)\rangle_{ab\vert c}$ near the lower boundary, both away from and close to the switching point $\zeta_1$. 

An important ingredient in our discussion is the average of the stress tensor in the strip geometry, given by
\begin{eqnarray} \label{stressstrip}
\left\langle T(z) \right\rangle_{ab\vert c}^{(\zeta_1)} = \left({\pi \over {\cal W}} \right)^2 \, \tau (\tilde{z}) \, , 
\quad \tau (\tilde{z}) = t_{ac} -{\hat{c} \over 24} + {t_{ab} \over (1-e^{-\tilde{z}})^2} + {t_{bc} - t_{ac}-t_{ab} \over 1-e^{-\tilde{z}}}\,
\end{eqnarray}
 for an arbitrary two-dimensional critical system. Here $\tilde{z} \equiv \pi (z -\zeta_1)/{\cal W}$, and $\hat{c}$ is the central charge of the system in the conformal classification \cite{BPZ,CardyD-L}, which equals ${1\over 2}$ for the Ising model. Expression~(\ref{stressstrip}) follows from the conformal mapping  $w(z)=\exp (\pi z / {\cal W})$ of the strip, with switching point $\zeta_1$, onto the upper half $w$ plane with two switches, from $c$ to $a$ at $w=0$ and from $a$ to $b$ at $w=\exp(\pi \zeta_1 / {\cal W})$. Combining this mapping with the average stress tensor (\ref{Tabc}) in the $w$ plane and the transformation property (\ref{Ttransform}) of the stress tensor leads to Eq.~(\ref{stressstrip}). 
\subsubsection{Expansion away from the switching point}
Averaging expansion (\ref{BOE}) in the $ab\vert c$ strip geometry and in the $ab$ half-plane geometry, subtracting the two averages, and substituting the corresponding stress tensors (\ref{Tab}) and (\ref{stressstrip}), we obtain
\begin{eqnarray} \label{MBOEwallaway}
\langle \phi(x,y)\rangle_{ab\vert c}^{(\zeta_1)}-\langle \phi(x,y)\rangle _{ab}^{(\zeta_1)}  \to \mu_h^{(\phi)}\, y^{-x_\phi}\left({\pi y\over{\cal W}}\right)^2 \, \left[ \tau (\tilde{x}) - {t_{ab}\over \tilde{x}^2}\right] 
\end{eqnarray}
for the distant-wall correction. Here $h=a$ and $h=b$ for $x< \zeta_1$ and $x> \zeta_1$, respectively. The asymptotic form (\ref{MBOEwallaway}) holds for $y$ much smaller than $|x-\zeta_1|$ and ${\cal W}$, but with no restriction on the scaling variable $\tilde{x}=\pi (x -\zeta_1)/{\cal W}$. In the limit $\tilde{x} \to -\infty$, Eqs.~(\ref{stressstrip}) and (\ref{MBOEwallaway}) reproduce the distant-wall correction \cite{TWBX,Cardydistantwall}, 
\begin{equation}
\langle \phi(x,y) \rangle_{a\vert c}-\langle \phi(x,y) \rangle _{a} \to {4x_\phi\over \hat{c}}\left({\hat{c}\over 24}-t_{ac}\right)
\langle\phi(x,y)\rangle_a \left({\pi y\over{\cal W}}\right)^2\,,\label{FdGcorrection}
\end{equation}
to the profile $\langle \phi(x,y) \rangle_{a\vert c}$ in a strip with uniform boundary conditions $a$ and $c$ on the edges. Here $\langle\phi(x,y)\rangle _a\propto y^{-x_\phi}$ is the profile in the half plane with boundary condition $a$, and we have used Eq.~(\ref{relationmuhphi}).  For $\tilde{x}\to\infty$, the corresponding result for $b\vert c$ boundaries is obtained. For $|\tilde{x}| \ll 1$, Eq.~ (\ref{MBOEwallaway}) yields
\begin{equation}
\langle \phi(x,y) \rangle_{ab\vert c}^{(\zeta_1)}-\langle \phi(x,y) \rangle _{ab} ^{(\zeta_1)}\to \mu_h^{(\phi)}\, y^{-x_\phi}\,{\pi(t_{bc}-t_{ac}) y^2\over (z-\zeta_1){\cal W}}\,,\label{correctionnearzeta1}
\end{equation} 
to leading order in the small quantities $y/|x-\zeta_1|$ and $y/{\cal W}$. 

According to Eqs.~(\ref{FdGcorrection}) and (\ref{correctionnearzeta1}), the distant-wall correction to the profile of $\phi$ falls off with increasing distance as ${\cal W}^{-2}$ for homogeneous boundaries and as ${\cal W}^{-1}$ near the switching point of $ab\vert c$ boundaries. 
The entire, smooth crossover between these two limiting cases is described by Eqs.~(\ref{stressstrip}) and (\ref{MBOEwallaway}).

\subsubsection{Expansion at the switching point}
The leading distant-wall correction to the profile in the neighborhood $\vert z-\zeta_1\vert \ll{\cal W}$ of the switching point follows from averaging the boundary-operator expansion (\ref{MBOE}) in the strip geometry, which yields
\begin{eqnarray} \label{MBOEnearzeta1}
\left\langle \phi(x,y) \right\rangle_{ab\vert c}^{(\zeta_1)}-\left\langle \phi(x,y) \right\rangle _{ab}^{(\zeta_1)}  \to 
F_{ab}^{(\phi)}(x-\zeta_1 , y) \, \left\langle\Upsilon({\zeta_1})\right\rangle_{ab\vert c}^{(\zeta_1)}\, .
\end{eqnarray}
Here the second term on the left-hand side  and the factor $F_{ab}^{(\phi)}(x-\zeta_1 , y)$ on the right are the same as in the half-plane  geometry. On the right only the second factor $\left\langle\Upsilon({\zeta_1})\right\rangle_{ab\vert c}^{(\zeta_1)}$ depends on the upper boundary. 

The explicit form of $\left\langle\Upsilon({\zeta_1})\right\rangle_{ab\vert c}^{(\zeta_1)}$ follows from setting $\phi=T$ in Eq.~(\ref{MBOEnearzeta1}), substituting the average stress tensors (\ref{Tab}) and (\ref{stressstrip}) on the left-hand side, and then expanding the left-hand side to leading non-vanishing order in $\tilde{z}$. Substituting expression (\ref{FabT}) for $F_{ab}^{(T)}$ on the right-hand side and solving for 
$\langle\Upsilon({\zeta_1})\rangle_{ab\vert c}^{(\zeta_1)}\,$, we obtain
\begin{eqnarray} \label{Upswall}
\langle\Upsilon({\zeta_1})\rangle_{ab\vert c}^{(\zeta_1)} =  {\pi(t_{bc} - t_{ac})\over{\cal W}}\, .
\end{eqnarray}

Thus, the distant-wall correction to the profile of $\phi$, where $\phi$ is either a primary operator or $T$, near the switching point is given by Eqs.~(\ref{MBOEnearzeta1}) and (\ref{Upswall}), together with the expressions for $F_{ab}^{(\phi)}$ in Eqs.~(\ref{FabT}) and (\ref{generalresult2text}), or, for the Ising model, in Eqs.~(\ref{dsigandepsabdzeta1}) and (\ref{Fabphi2}).

The assumption $|z-\zeta_1|\ll {\cal W}$ made in this subsection and the assumption $y\ll |x-\zeta_1|$, $y\ll {\cal W}$ of the preceding subsection are both satisfied if $y\ll |x-\zeta_1|\ll{\cal W}$. Thus,  for $y\ll |x-\zeta_1|\ll{\cal W}$, the distant-wall predictions (\ref{MBOEnearzeta1}) and (\ref{correctionnearzeta1}) of the boundary-operator expansions at and away from the switching point should coincide. Substituting Eq.~(\ref{Upswall}) and the asymptotic form (\ref{Fhom}) of $F_{ab}^{(\phi)}$ in Eq.~(\ref{MBOEnearzeta1}), we see that this is indeed the case.

In the Ising model with $c=f$ and $ab=+-$ or $ab=-+$, $\langle \sigma \rangle$ is an odd function of $x-\zeta_1$, while $\langle \epsilon \rangle$ is even. That the corresponding $F_{ab}^{(\sigma)}$ and $F_{ab}^{(\epsilon)}$, given by Eqs.~(\ref{dsigandepsabdzeta1}) and (\ref{Fabphi2}) are even and odd, respectively, is inconsistent with Eq. (\ref{MBOEnearzeta1}) unless $\langle \Upsilon \rangle_{ab\vert c}$ vanishes. According to Eq.~(\ref{Upswall}), $\langle \Upsilon \rangle_{ab\vert c}$ does indeed vanish in these two cases, and we conclude that the leading distant-wall correction is of higher order. 

For all other  $ab\vert c$ with $a \neq b$, the expression for $\langle \Upsilon \rangle_{ab\vert c}$ in (\ref{Upswall}) is non-vanishing, and it is instructive to compare the signs of the predicted distant-wall corrections to $\langle\sigma\rangle$ and $\langle\epsilon\rangle$ with one's intuitive expectation.

Finally, we have also confirmed the predictions  (\ref{MBOEwallaway}), (\ref{MBOEnearzeta1}), and (\ref{Upswall}) of the two boundary-operator expansions for the Ising model by comparison with exact expressions for $\langle\sigma\rangle_{ab\vert c}$ and $\langle\epsilon\rangle_{ab\vert c}\,$, derived from half-plane results with the conformal mapping mapping onto the strip discussed below Eq. (\ref{stressstrip}).}

\section{Concluding remarks}\label{concludingremarks}

In the first half of this paper (see Sec.~\ref{IsingConformal}), the semi-infinite critical Ising model with mixed boundary conditions $+f+f+\dots$ and $-f+$ is analyzed with  conformal-invariance methods. Exact expressions for the one and two-point averages $\langle\sigma\rangle$,  $\langle\epsilon\rangle$, $\langle T\rangle$, $\langle\sigma_1\sigma_2\rangle$, $\langle\epsilon_1\epsilon_2\rangle$, $\langle\sigma_1\epsilon_2\rangle$ are derived. The additional averages  $\langle T_1T_2\rangle$, $\langle T_1\sigma_2\rangle$, $\langle T_1\epsilon_2\rangle$, $\langle T_1\sigma_2\sigma_3\rangle$, etc. are readily obtained by substituting these results into expressions (\ref{TTabcdots}),(\ref{TTbarabcdots})  for $\langle T_1T_2\rangle$ and the conformal Ward identity, e.g. Eq.~(\ref{GGWI}). The results of Sec.~\ref{IsingConformal} complement the predictions for $+-+-\dots$ boundary conditions in Ref.~\cite{TWBG2}.

In our approach we profit from the fact that the amplitude $t_{+f}$ of the stress tensor $\langle T\rangle_{ab}$ and the scaling indices $\Delta_\sigma$ and $\Delta_\mu$ of the spin and disorder operators all have the same value ${1\over 16}$. Consequently, all the multi-spin averages $\langle\sigma_1\sigma_2\dots\sigma_n\dots\rangle$ with $+f+f+\dots$ and $-f+$ boundary conditions can be expressed in terms of the known solutions (\ref{Gnsigma}) of the bulk conformal differential equations for $\Delta={1\over 16}$.  To calculate averages involving $\epsilon$ from the multi-spin averages, we used the operator product expansion (\ref{OPEsigsig}) for $\sigma\sigma$.
   
In future work we plan to consider other two-dimensional critical systems, such as the $Q$-state Potts and ${\rm O}(N)$ models, with mixed boundary conditions. The Potts profiles $\langle{\bf\sigma}\rangle_{ab}$ and $\langle\epsilon\rangle_{ab}$ for  general $Q$ have already been determined \cite{TWBX}.

The second half of this paper (see Sec.~\ref{secMBOE}) is devoted to boundary-operator expansions in two-dimensional critical systems with mixed boundary conditions and is not limited to the Ising model. Two types of expansions, at and away from switching points of the boundary condition, are considered. Apart from the case of the order parameter near a free boundary, the leading boundary operator in the expansion away from a switching point is the complex stress tensor $T(x)$ at the surface, which has scaling dimension 2.  In contrast, in the expansion at a switching point $\zeta_1$, the leading boundary operator $\Upsilon(\zeta_1)$ has scaling dimension 1. We demonstrate the utility of the two expansions in predicting the asymptotic behavior of many-point averages and distant wall corrections to one-point averages in the strip geometry. 

Finally, we point out  the utility of boundary-operator expansions, not only at switching points of the boundary condition, but also at points where the boundary bends abruptly, for example at the tip of a wedge or needle. The  asymptotic behavior near the tip of a semi-infinite needle with a single boundary condition, immersed in a two-dimensional critical fluid, is analyzed with the help of a boundary-operator expansion in Appendix \ref{OEneedle}.

\appendix

\section{Check of the boundary-operator expansion away from the switching point of a $+-$ boundary}\label{appendixcheckaway}
Here we confirm that the exact two-point average  $\langle\sigma\sigma \rangle_{+-}^{\rm cum}$, given in Eqs.~(4.1) and (4.3) of Ref.~\cite{TWBX}, has the asymptotic behavior (\ref{twoaway}) for $y \to 0$, $x\ne\zeta_1$ predicted by the boundary operator expansion away from a switching point. 
Expressed in terms of the positions $(x,y)$ and  $(X,Y)$ of the two spin operators and the angles $\vartheta$ and $\Theta$ defined in Eq.~(\ref{rthetaRTheta}), the exact result takes the form
\begin{eqnarray}
&&\langle\sigma\sigma\rangle_{+-}^{\rm cum}=\langle\sigma\sigma\rangle_{-+}^{\rm cum} = \langle\sigma\rangle_{_ +}^{(y)}\langle\sigma\rangle_{_+}^{(Y)}\nonumber\\
&&\qquad\quad\times\;{1\over\sqrt 2}\left\{\left[\left(u+u^{-1}\right)^{1/2}-\sqrt 2\,\right]\cos\vartheta\cos\Theta+{u-u^{-1}\over\left(u+u^{-1}\right)^{1/2}}\,\sin\vartheta\sin\Theta\right\}\,,\label{checkkumuplusminus}\\
&&u=\left[1+{{4yY}\over (x-X)^2+(y-Y)^2}\right]^{1/4}.\label{defineu}
\end{eqnarray}
Expanding Eqs.~(\ref{checkkumuplusminus}) and (\ref{defineu})
for $y$ much smaller than $|x-\zeta_1|$ and $\left[(X-x)^2+Y^2\right]^{1/2}$ leads to
\begin{eqnarray} \label{twoaway'}
&&\langle\sigma\sigma\rangle_{+-}^{\rm cum}\;\to\;\langle\sigma\rangle_{_+}^{(y)}\langle\sigma\rangle_{_+}^{(Y)} \nonumber\\
&&\qquad\quad\times\;{ {\rm sgn}(x-\zeta_1)\over\sqrt{(X-\zeta_1)^2 +Y^2}}\;{(y Y)^2 \over\left[(X-x)^2 +Y^2\right]}\left[{1 \over 4}\, {X-\zeta_1 \over  (X-x)^2 +Y^2} + {1 \over x-\zeta_1} \right]\,,\label{lefthandside}
\end{eqnarray}
to leading, non-vanishing order. 

Continuing our check of the asymptotic form (\ref{twoaway}) for $\langle\sigma\sigma\rangle_{+-}^{\rm cum}$,  we next evaluate the right-hand side of Eq.~(\ref{twoaway}), using the Ward Identity (\ref{GGWI}) with $\Phi = \sigma$ and $z=x$, and substituting the exact result $\langle\sigma\rangle_{+-}=-\langle\sigma\rangle_{_{+}}^{(Y)}\cos{\Theta}\;$ for $\langle \Phi(X,Y)\rangle_{ab\{c\}}$. For the $+-$ boundary condition, $\sum_{j}$ only contains the term with $j=1$. Expressing $\cos\Theta$ in terms of $Z$, $\bar Z$, and $\zeta_1$ with the help of Eq.~(\ref{rthetaRTheta}), evaluating the right-hand side of the Ward identity explicitly, and substituting the result on the right-hand side of Eq.~(\ref{twoaway}) leads to the same result as in Eq.~(\ref{lefthandside}). 
This confirms that the asymptotic behavior of the exact two-point average $\langle\sigma\sigma\rangle_{+-}$  for $y\to 0$, $x\neq\zeta_1$ is in complete agreement with the prediction (\ref{twoaway}) of the boundary operator expansion away from a switching point.

\section{Derivation of the relation (\ref{generalresult2text}) between $F_{ab}^{(\phi)}$ and $\partial_{\zeta_1}\langle\phi\rangle_{ab}$}\label{appendixderivation}

For consistency with scaling, the one-point average $\langle\phi\rangle_{ba}$ of a primary operator $\phi$ in a two-dimensional critical system with 
a $ba$ boundary must have the form
\begin{equation}
\langle\phi\rangle_{ba}=y^{-x_\phi}H_{ba}(\theta_1)\,,\;\;\; H_{ab}(\theta_1)=H_{ba}(\pi-\theta_1)\,,\;\;\; \theta_1={\rm arg}(z-\zeta_1)\,.\label{defineH}
\end{equation}

As mentioned in connection with Eqs.~(\ref{Taba}) and (\ref{sigpfp}), $\langle\phi\rangle_{aba}$ follows from $\langle\phi\rangle_{ba}$ under the conformal mapping  (\ref{abtoaba}).
The end effect of the mapping  is to replace $\theta_1$ in Eq.~(\ref{defineH}) with
\begin{equation}
\gamma_{_{1,1}}={\rm arg}\left[(z-\zeta_2)/(z-\zeta_1)\right]=\theta_2-\theta_1\,,
\end{equation}
so that
\begin{eqnarray}
\langle\phi\rangle_{aba}=y^{-x_\phi}H_{ba}(\gamma_{_{1,1}})=y^{-x_\phi}H_{ab}(\pi-\theta_2+\theta_1)\,.
\end{eqnarray}
For $z$ close to $\zeta_1$, 
\begin{equation}
\theta_2={\rm arg}(z-\zeta_2)=\arctan\left({y\over x-\zeta_2}\right)\to \pi-{y\over\zeta_2-\zeta_1}\,.
\end{equation}
Thus,
\begin{eqnarray}
\langle\phi\rangle_{aba}-\langle\phi\rangle_{ab}&\to& y^{-x_\phi}\left[H_{ab}\left(\theta_1+{y\over\zeta_2-\zeta_1}\right)-H_{ab}(\theta_1)\right]\nonumber\\[3mm]
&\to& y^{-x_\phi}\,H_{ab}'(\theta_1)\,{y\over \zeta_2-\zeta_1}={(x-\zeta_1)^2+y^2\over\zeta_2-\zeta_1}\,\partial_{\zeta_1}\langle\phi\rangle_{ab}\,,\label{generalresult0}
\end{eqnarray}
where, in obtaining the rightmost expression, we have used $\theta_1=\arctan\left[y/(x-\zeta_1)\right]$.

Comparing this result with Eq.~(\ref{MBOEone}), we conclude that
\begin{equation}
F_{ab}(x-\zeta_1,y)\langle\Upsilon(\zeta_1)\rangle_{aba}^{(\zeta_1,\zeta_2)}={|z-\zeta_1|^2\over\zeta_2-\zeta_1}\,\partial_{\zeta_1}\langle\phi\rangle_{ab}\,.\label{generalresult1}
\end{equation}
Substituting expression (\ref{avupsaba}) for $\langle\Upsilon(\zeta_1)\rangle_{aba}^{(\zeta_1,\zeta_2)}$ in Eq.~(\ref{generalresult1}) leads to the relation (\ref{generalresult2text}) between $F_{ab}^{(\phi)}$ and $\partial_{\zeta_1}\langle\phi\rangle_{ab}$ that we set out to prove.

Just above Eq.~(\ref{FabT}) we stated that the scaling dimension 
of $\Upsilon$ equals 1, not just for the Ising model, but for other two-dimensional critical systems as well. This follows from Eq. (\ref{generalresult1}).
Recalling that $F_{ab}^{(\phi)}$ depends on $x$ and $y$ but not on $\zeta_2$, while $\langle\Upsilon(\zeta_1)\rangle_{aba}$ depends on $\zeta_2$ but not on $x$ and $y$, we conclude that $\langle\Upsilon(\zeta_1)\rangle_{aba}^{(\zeta_1,\zeta_2)}\propto (\zeta_2-\zeta_1)^{-1}$. Thus, the scaling dimensions of $\Upsilon$ and $F_{ab}^{(\phi)}$ are 1 and  $x_\phi -1$ respectively.

No properties specific to the Ising model were used in the steps leading to Eqs.~(\ref{generalresult1}) and (\ref{generalresult2text}). These relations are expected to hold for primary operators $\phi$ in other two-dimensional critical systems, such as the $Q$-state Potts model, for which some of the 
$\langle\phi\rangle_{ab}$ are known explicitly \cite{TWBX}.

\section{Relation of $\langle\Upsilon(\zeta_j)\rangle_{abcd...}$ to the free energy}\label{relationUpsfreeenergy}
The free energy per $k_B T$ of a two-dimensional critical system in the upper half plane with area $A$, boundary extending from $-{1\over 2}L$ to ${1\over 2}L$ along the $x$ axis, and $abc$ boundary conditions is given by
\begin{eqnarray} \label{F}
F= A f^{({\rm bulk})} + (\zeta_1+{\textstyle{1\over 2}}L) f_{a}^{({\rm s})} + (\zeta_2 -\zeta_1) f_{b}^{({\rm s})} +({\textstyle{1\over 2}}L -\zeta_2) f_{c}^{({\rm s})} + {\cal F} 
\end{eqnarray}
for large $L$.
Here $f^{({\rm bulk})}$ is the bulk free energy per unit area, and $f_{a}^{({\rm s})}$, $f_{b}^{({\rm s})}$, and $f_{c}^{({\rm s})}$ are the surface free energies per unit length for uniform boundaries $a$, $b$, $c$. The final term ${\cal F}$ is the free energy of interaction between the boundary switches at $\zeta_1$ and $\zeta_2$, which has the universal form \cite{CardyD-L,Cardyscp,EETWB}
\begin{eqnarray} \label{partialF}
\partial_{\zeta_2 - \zeta_1} {\cal F} &=& - \int\limits_{0}^{+ \infty} d y \, \langle T_{xx} (x_0 ,y) \rangle_{abc} = \int\limits_{\cal C} {dz \over 2 \pi i}\, \langle T (z) \rangle_{abc} = \nonumber \\
&=& \langle \Upsilon(\zeta_1) \rangle_{abc} = - \langle \Upsilon(\zeta_2) \rangle_{abc} = {t_{ab}+t_{bc}-t_{ac} \over \zeta_2 - \zeta_1}\,,
\end{eqnarray}
where $\zeta_1 < x_0 <\zeta_2$ and the integration path $\cal C$ extends from  from $y=-\infty$ to $y=+\infty$ along a vertical line that crosses the $x$ axis at $x_0$. Here we have used the relation $T_{xx}(x,y)= - \left[T(z) + \bar T(\bar z)\right]/2 \pi$ between the Cartesian and complex stress tensors \cite{CardyD-L}, with $\bar T(\bar z)=T(\bar z)$ in the half-plane geometry \cite{Cardyscp}. In going from line 1 to line 2, we evaluated the  integral using Cauchy's theorem, after closing the integration path $\cal C$ with an infinite left or right semicircle, both with $\langle T(z)\rangle_{abc}$ taken from Eq.~(\ref{Tabc}) and formed from boundary-operator expansions (\ref{expandT1}) and (\ref{expandT2}) for left and right semicircles, respectively. 
Since $f^{({\rm bulk})}$ and $f_{a}^{({\rm s})},$..., $f_{c}^{({\rm s})}$ in Eq.~(\ref{F}) are independent of the switching points,
\begin{eqnarray}
&&\partial_{\zeta_1} F = f_{a}^{({\rm s})}- f_{b}^{({\rm s})} - \langle \Upsilon(\zeta_1) \rangle_{abc}\,,  \label{partialF1}\\
&&\partial_{\zeta_2} F = f_{b}^{({\rm s})}- f_{c}^{({\rm s})} - \langle \Upsilon(\zeta_2) \rangle_{abc}\,,\label{partialF2}
\end{eqnarray}
for fixed $\zeta_2$ and $\zeta_1$, respectively.

Equations (\ref{partialF1}) and (\ref{partialF2}) provide a direct physical interpretation of the universal quantities $\langle \Upsilon(\zeta_1) \rangle_{abc}$ and $ \langle \Upsilon(\zeta_2) \rangle_{abc}$ in the boundary-operator expansion. They represent the universal, fluctuation-induced or Casimir part of the force on switching point 1 due to switching point 2 and the equal and opposite force on switching point 2, respectively. The contributions of the non-universal quantities $f_{a}^{({\rm s})}$, $f_{b}^{({\rm s})}$, $f_{c}^{({\rm s})}$ to the attraction or repulsion depend on microscopic details. These contributions are independent of $\zeta_1$ and $\zeta_2$, unlike the universal contributions $\langle \Upsilon(\zeta_1) \rangle_{abc}$ and $ \langle \Upsilon(\zeta_2) \rangle_{abc}$, which vary as $(\zeta_2-\zeta_1)^{-1}$.


According  to Eq.~(\ref{partialF}) and the values of $t_{+f}={1\over 16}$, $t_{+-}={1\over 2}$ for the Ising model, on decreasing the separation $\zeta_2-\zeta_1$, the universal quantity ${\cal F}$ decreases for boundary conditions $aba$ and $f+-$ but increases for $+f-$. This is plausible, since for $aba$ and $\zeta_2-\zeta_1\searrow 0\,$ the energetically-advantageous uniform boundary is approached , for $f+-$ the energy-costly $+-$ switch is removed, and for $+f-$ it is created. 

For $abcd...$ boundary conditions Eqs.~(\ref{partialF1}) and (\ref{partialF2}) are replaced by
\begin{equation}
\partial_{\zeta_j} F = f_{j}^{({\rm s})}- f_{j+1}^{({\rm s})} - \langle \Upsilon(\zeta_j) \rangle_{abcd...}\,,\label{partialFzetaj}
\end{equation}
To derive this relation, we begin with the same integral as in Eq.~(\ref{partialF}), but with crossing point $x_0$ between $\zeta_{j-1}$ and $\zeta_j$, and subtract from it the same integral, but with crossing point between $\zeta_{j}$ and $\zeta_{j+1}$. In this way the value of $\zeta_j$ is increased, while all the other $\zeta$'s are kept fixed. Combining the two integrals into a single integral with a path that encircles $\zeta_j$ clockwise, forming $\langle T(z)\rangle_{abcd...}$ from the boundary-operator expansion analogous to (\ref{expandT2}), and using Cauchy's theorem, we obtain $\partial_{\zeta_j} {\cal F} = - \langle \Upsilon(\zeta_j) \rangle_{abcde...}$, which leads with straightforward steps to Eq.~(\ref{partialFzetaj}).

\section{Two-point correlations of $\Upsilon$}\label{appendixUpsUps}

By combining the boundary operator expansion at a switching point $\zeta_j$ other than $\zeta_1$ with Eq. (\ref{UpsD}), the two-point function $\langle \Upsilon (\zeta_1) \Upsilon(\zeta_j) \rangle_{abcd...}$ can be calculated. Here this is illustrated in the simplest case  $\langle \Upsilon (\zeta_1)\Upsilon(\zeta_2) \rangle_{abc}$.

For $z$ near the $ab$ switching point $\zeta_1$, expansion (\ref{MBOE}), for $\phi=T$, can be expressed as
\begin{equation}
T(z)\to {t_{ab}\over (z-\zeta_1)^2}+{1\over z-\zeta_1}\Upsilon(\zeta_1)\,,\label{expandT1}
\end{equation}
with the help Eqs.~(\ref{Tab}) and (\ref{FabT}). Similarly for $z$ near the $bc$ switching point $\zeta_2$,
\begin{equation}
T(z)\to {t_{bc}\over (z-\zeta_2)^2}+{1\over z-\zeta_2}\Upsilon(\zeta_2)\,.\label{expandT2}
\end{equation}
Averaging Eq. (\ref{expandT1}) with $abc$ boundary conditions, substituting expression Eq.~(\ref{Tabc}) for $\langle T(z)\rangle_{abc}$, and equating the  
leading terms for $|z-\zeta_1|\ll \zeta_2-\zeta_1$ on the left and right-hand sides leads to expression (\ref{avupsaba}) for $\langle \Upsilon(\zeta_1) \rangle_{abc}$. From an analogous calculation based on Eq.~(\ref{expandT2}), we conclude
\begin{eqnarray} \label{Ups2}
\langle \Upsilon(\zeta_2) \rangle_{abc} = - \langle \Upsilon(\zeta_1) \rangle_{abc} ={t_{ac}-t_{ab}-t_{bc}\over \zeta_2-\zeta_1}\,.\label{Ups12}
\end{eqnarray}

To calculate $\langle\Upsilon_1\Upsilon_2\rangle_{abc}$, we set $ab\{ c\}=abc$  and replace $\Phi(X,Y)$ by $T(z)$ in Eq.~(\ref{UpsD}). Then, on substituting expansion (\ref{expandT2}) for $T(z)$ on the left-hand side and expression (\ref{Tabc}) for $\langle T \rangle_{abc}$ on the right-hand side, picking out the dominant terms for $z$ near $\zeta_2$, and using Eq.~(\ref{Ups12}), we obtain
\begin{equation} 
\langle \Upsilon ({\zeta_1}) \Upsilon ({\zeta_2}) \rangle_{abc}^{\rm cum} =  {t_{ac}-t_{ab}-t_{bc}\over (\zeta_2 -\zeta_1)^2 } \, .\label{UpsUps} 
\end{equation}
for the cumulant or connected part of the two-point average.

This result also follows from the exact expression for $\langle T(z_0)T(z)\rangle_{abc}$ in Eq.~(\ref{TTcumabc}) on substituting expansions (\ref{expandT1}) and (\ref{expandT2}) for $T(z_0)$ and $T(z)$, respectively, and identifying the dominant terms for for $z_0$ near $\zeta_1$ and $z$ near $\zeta_2$.

\section{Two-point averages of the stress tensor}\label{appendixTT}
For an arbitrary conformally-invariant, semi-infinite critical system with central charge $\hat{c}$ \cite{BPZ,CardyD-L} and with mixed boundary conditions, the components $T(z)$ and $\bar T(\bar z)$  of the complex stress tensor satisfy the identities. 
\begin{eqnarray}
&&\langle T(z_0) T(z) \rangle^{\rm cum} _{abc...}=  {\hat{c}/2 \over (z_0- z)^4}\nonumber \\ 
&&\qquad\qquad +\left[ {2 \over (z_0 -z)^2} +{1 \over z_0 -z}\, \partial_z + \sum_j {1 \over z_0 -\zeta_j}\,\partial_{\zeta_j}\right] \langle T(z) \rangle_{abc...}\,, \label{TTabcdots}\\[2mm]
&&\langle T(z_0) \bar{T}(\bar{z}) \rangle^{\rm cum}_{abc...} ={\hat{c}/2 \over (z_0- \bar{z})^4}\nonumber \\ 
&&\qquad\qquad +\left[{2 \over (z_0 -\bar{z})^2} +{1 \over z_0 -\bar{z}}\,\partial_{\bar{z}} +\sum_j {1 \over z_0 -\zeta_j}\, \partial_{\zeta_j}\right] \langle \bar{T}(\bar{z}) \rangle_{abc...}\,.\label{TTbarabcdots}
\end{eqnarray}
These relations follow from the same steps as in Cardy's derivation \cite{Cardyscp} of the conformal Ward identity in the half-plane geometry and its extension to mixed boundary conditions \cite{TWBX}, except that the conformal transformation $\phi(z,\bar z)\to|w'(z)|^{x_\phi}\phi(w,\bar w)$ for primary operators is replaced by \cite{BPZ,CardyD-L}
\begin{eqnarray}
&&T(z)\to w'(z)^2\, T(w)+{\textstyle{1\over 12}}\,\hat{c}\,\{w,z\}\,,\quad\{w,z\}\equiv{w'''(z)\over w'(z)}-{\textstyle{3\over 2}}\left[{w''(z)\over w'(z)}\right]^2\,,\label{Ttransform}
\end{eqnarray}
and its conjugate. Here $\{w,z\}=-w'(z)^2\{z,w\}$ is the Schwarzian derivative, already encountered in Eq.~(\ref{tautau}).
 
The identities (\ref{TTabcdots}) and  (\ref{TTbarabcdots}) may also be derived by substituting the operator-product expansion
\begin{equation}
\epsilon_1\epsilon_2\to\vert z_{12}\vert^{-2x_\epsilon}\left\{1+{x_\epsilon\over\hat{c}}\left[z_{12}^2\,T(z)+\bar z_{12}^2\,\bar T(\bar z)\right]+...\right\}\,,\quad z={\textstyle{1\over 2}}(z_1+z_2)\,,\label{OPEepsepsgeneral}
\end{equation}
in the conformal Ward identity relating $\langle T(z_0)\epsilon_1\epsilon_2\rangle_{abc..}$ and 
$\langle\epsilon_1\epsilon_2\rangle_{abc..}$
and equating the terms proportional $z_{12}^2/|z_{12}|^{2x_{\epsilon}}$ on both sides, likewise for the terms proportional to ${\bar z}_{12}^2/|z_{12}|^{2x_{\epsilon}}$. The operator-product expansion (\ref{OPEepsepsgeneral}) is established for general $\hat{c}$ in Ref.~\cite{EE}. For $\hat{c}={1\over 2}$ and $x_\epsilon=1$, corresponding to the Ising model, it reduces to the expansion (\ref{OPEepseps}).

The explicit form of $\langle T(z_0) T(z) \rangle^{\rm cum}$ for a semi-infinite critical system with an $ab$ boundary is obtained by substituting expression (\ref{Tab}) for $\langle T(z) \rangle_{ab}$ in Eq. (\ref{TTabcdots}). This leads to
\begin{eqnarray} 
\langle T(z_0)T(z) \rangle^{{\rm cum}}_{ab}= {\hat{c}/2 \over (z_0- z)^4}+{2 t_{ab}\over (z_0 -z)^2} {1 \over (z_0 -\zeta_1) (z-\zeta_1)}\,.\label{TTcumab}
\end{eqnarray}
For an $abc$ boundary, Eqs.~(\ref{Tabc}) and (\ref{TTabcdots}) yield
\begin{eqnarray} 
&&\langle T(z_0)T(z) \rangle^{{\rm cum}}_{abc}= {\hat{c}/2\over (z_0- z)^4}+{1\over (z_0 -z)^2}\bigg[ {2t_{ab} \over (z_0 -\zeta_1) (z-\zeta_1)}
+ {2t_{bc}\over (z_0 -\zeta_2) (z-\zeta_2)} \nonumber\\[2mm]
&&\qquad\qquad\qquad\qquad+ \left(t_{ac}-t_{ab}-t_{bc}\right) {2(z_0 z + \zeta_1 \zeta_2)-(z_0 +z) (\zeta_1 + \zeta_2) \over (z_0 -\zeta_1) (z_0-\zeta_2)(z -\zeta_1) (z-\zeta_2)} \bigg]\,.\label{TTcumabc}
\end{eqnarray}
Although the invariance under exchange of $z_0$ and $z$ is not immediately apparent on the right-hand side of Eq.~(\ref{TTabcdots}), it is obvious in Eqs.~(\ref{TTcumab}) and (\ref{TTcumabc}).

\section{Check of the boundary-operator expansion at a switching point}\label{appendixcheckat}
\subsection{Comparison of results for $\langle\sigma\sigma\rangle_{+-}^{\rm cum}$}
The check begins with the exact result for $\langle\sigma\sigma\rangle_{+-}^{\rm cum}$ in Eqs.~(\ref{checkkumuplusminus}) and (\ref{defineu}). In terms of the polar coordinates defined in Eq.~(\ref{rthetaRTheta}),
\begin{equation}
u=\left[1+{4(r/R)\sin\vartheta\sin\Theta\over 1+2(r/R)\cos(\vartheta-\Theta)+(r/R)^2}\right]^{1/4}.\label{polaru}
\end{equation} 

For $(x,y)$ close to the switching point $(\zeta_1,0)$, the ratio $r/R$ is a small quantity, and to first order,
\begin{equation}
u=1+(\sin\vartheta\sin\Theta)\,{r\over R}+{\rm O}\left((r/R)^2\right),\label{uexpand}                                                                                                                                                                                                                                                                                                                                                                                                                                                                                                                                                                           
\end{equation}
Substituting Eq.~(\ref{uexpand}) in Eq.~(\ref{checkkumuplusminus}) and expanding the curly bracket in Eq.~(\ref{checkkumuplusminus}) to first order in $r/R$ leads to
\begin{eqnarray}
&&\langle\sigma\sigma\rangle_{+-}^{\rm cum}\;\to\;\langle\sigma\rangle_{_+}^{(y)}\langle\sigma\rangle_{_+}^{(Y)}\left(\sin\vartheta\sin\Theta\right)^2{r\over R}=F_{+-}^{(\sigma)}\times\partial_{\zeta_1}\langle\sigma\rangle_{+-}\,,\label{expandkumuplusminus}
\end{eqnarray}
where, in going from the first expression on the right-hand side of Eq. (\ref{expandkumuplusminus}) to the second expression, we have made use of
Eqs.~(\ref{dsigandepsabdzeta1}) and (\ref{Fabphi2}).
Comparing Eqs.~(\ref{kumu}) and (\ref{expandkumuplusminus}), we see that the asymptotic behavior of the exact expression for 
$\langle\sigma\sigma\rangle_{+-}^{\rm cum}$ is in complete agreement with the prediction of the operator expansion for $\phi=\Phi=\sigma$ and 
$ab=+-$. Proceeding in this way, we have confirmed the consistency for the other combinations of $\phi$ and $\Phi$ equal to $\sigma$ and $\epsilon$ and for $ab=+-$ and $+f$.

\subsection{Comparison of results for $\langle\sigma\rangle_{+f+}$ and $\langle\sigma\sigma\rangle_{+f+}^{\rm cum}$}
According to the exact results in Eqs.~(\ref{sigpfp}), (\ref{sigsigpfp}), and (\ref{sigandepsab}),
\begin{eqnarray}
&&\langle\sigma\rangle_{+f+}-\langle\sigma\rangle_{+f}=\langle\sigma\rangle_{_+}^{(y)}\left[\left(\cos{\textstyle{\gamma\over 2}}\right)^{1/2}
-\left(\sin{\textstyle{\vartheta\over 2}}\right)^{1/2}\right]\label{checkavphi}\\[4mm]
&&\langle\sigma\sigma\rangle_{+f+}^{\rm cum}=\langle\sigma\rangle_{_+}^{(y)}\langle\sigma\rangle_{_+}^{(Y)}\bigg\{{1\over \sqrt{2}}\left[{1\over\sqrt{\rho}}\cos\left(\textstyle{\gamma\over 2}-\textstyle{\Gamma\over 2}\right)+\sqrt{\rho}\cos\left(\textstyle{\gamma\over 2}+\textstyle{\Gamma\over 2}\right)\right]^{1/2}\nonumber\\ 
&&\qquad\qquad\qquad\qquad\qquad\qquad -
\Big[\cos{\textstyle{\gamma\over 2}}\cos{\textstyle{\Gamma\over 2}}\Big]^{1/2}\bigg\}.\label{checkkumu}
\end{eqnarray}
Here we have replaced the positions $(x_1,y_1)$ and $(x_2,y_2)$ by $(x,y)$ and $(X,Y)$, and $\gamma_{1,1}$ and $\gamma_{2,1}$ by $\gamma$ and $\Gamma$, respectively. In terms of Cartesian coordinates and the polar coordinates defined in Eq.~(\ref{rthetaRTheta}),
\begin{eqnarray}
&&\rho=\left[{(x-X)^2+(y-Y)^2\over (x-X)^2+(y+Y)^2}\right]^{1/2}=
\left[{1-2(r/R)\cos(\vartheta-\Theta)+(r/R)^2\over1-2(r/R)\cos(\vartheta+\Theta)+(r/R)^2}\right]^{1/2},\label{rhonew}\\[2mm]
&&\gamma={\rm arg}\left({ z-\zeta_2\over z-\zeta_1}\right)=\pi-\vartheta-\arctan\left[{(r/\zeta_{21})\sin\vartheta\over 
1-(r/\zeta_{21})\cos\vartheta}\right]\,.\label{definegamma}
\end{eqnarray}
Analogous expressions for $\Gamma$ are shown in Eqs.~(\ref{defineGamma}) and (\ref{trigGamma}).

For $z$ close to the switching point $\zeta_1$, the ratios $r/R$ and $r/\zeta_{21}$, are small quantities, and to first order,
\begin{eqnarray}
&&\rho=1-(2\sin\vartheta\sin\Theta)\,{r\over R}+{\rm O}\left((r/R)^2\right),\label{rhoexpand}\\[2mm]
&&\gamma=\pi-\vartheta-(\sin\vartheta)\,{r\over\zeta_{21}}+{\rm O}\left((r/\zeta_{21})^2\right)\,.\label{gammaexpand}                                                                                                                                                                                                                                                                                                                                                                                                                                                                                                                                                                                  
\end{eqnarray}
Substituting Eqs.~(\ref{rhoexpand}) and (\ref{gammaexpand}) and expanding the square bracket in Eq.~(\ref{checkavphi}) and the curly bracket in Eq.~(\ref{checkkumu}) to first order in $r/R$ and $r/\zeta_{21}$, we obtain
\begin{eqnarray}
&&\langle\sigma\rangle_{+f+}-\langle\sigma\rangle_{+f}\;\to\;
{1\over 2}\langle\sigma\rangle_{+}^{(y)}\left(\sin{\textstyle{\vartheta\over 2}}\right)^{1/2}\left(\cos{\textstyle{\vartheta\over2}}\right)^2{r\over\zeta_{21}}
=F_{+f}^{(\sigma)}\times\langle\Upsilon\rangle_{+f+}^{(\zeta_1,\zeta_2)}\,,\label{expandavphi}\\[2mm]
&&\langle\sigma\sigma\rangle_{+f+}^{\rm cum}\;\to\;\langle\sigma\rangle_{+}^{(y)}\langle\sigma\rangle_{+}^{(Y)}\left(\sin{\textstyle{\vartheta\over 2}}\right)^{1/2}\left(\cos{\textstyle{\vartheta\over2}}\right)^2\sin\Theta\,{\sin{\textstyle{\Gamma\over 2}}\over\sqrt{\cos{\textstyle{\Gamma\over 2}}}}\,{r\over R}\nonumber\\  &&\qquad\qquad\qquad\qquad\qquad =F_{+f}^{(\sigma)}\times\partial_{\zeta_1}\langle\sigma\rangle_{+f+}\,.\label{expandkumu}
\end{eqnarray}
In going from the first expression on the right-hand sides of Eqs.~(\ref{expandavphi}) and (\ref{expandkumu}) to the second expression on the right, we have used the expression for $\partial_{\zeta_1}\langle\sigma\rangle_{+f+}$ and $F_{+f}^{(\sigma)}$, obtained as described below Eq.~(\ref{dsigandepsabdzeta1}), and the relation  $\langle\Upsilon\rangle_{+f+}^{(\zeta_1,\zeta_2)}={1\over 8}(\zeta_2-\zeta_1)^{-1}$, which follows from Eq.~(\ref{avupsaba}), with $t_{+f}=t_{f+}={1\over 16}$ and $t_{++}=0$. The asymptotic behavior of the exact one and two-point averages, shown in Eqs. (\ref{expandavphi}) and (\ref{expandkumu}), is in complete agreement with the predictions (\ref{MBOEone}) and (\ref{kumu}) of the operator expansion for $\phi=\Phi=\sigma$ and $abc=+f+$. Proceeding in this way, we have confirmed the consistency for the other combinations of $\phi$ and $\Phi$ equal to $\sigma$ and $\epsilon$ and the $abc$ boundary conditions considered in Subsec.~\ref{abcboundaries}.

\section{Asymptotic behavior near the tip of a needle}\label{OEneedle}
Consider a semi-infinite needle in the full $z=x+iy=re^{i\vartheta}$ plane that extends from the origin along the positive real axis to $x=+ \infty$ and has boundary condition $a$ on both sides. Under the conformal mapping $w=z^{1/2}$ or, equivalently, $u=r^{1/2}\cos{\vartheta\over 2}$, $v=r^{1/2}\sin{\vartheta\over 2}$, the complex $z$ plane with this boundary condition is mapped onto the upper half $v>0$ of the complex $w=u+iv$ plane with boundary condition $a$ along the entire $u$ axis. We begin with the useful relations
\begin{eqnarray}
&&\langle\phi(u,v)\rangle_a=A_a^{(\phi)}v^{-x_\phi}\,,\quad\langle\phi(x,y)\rangle_{{\rm ndl},a}=A_a^{(\phi)}\left(2r\sin{\textstyle{\vartheta\over 2}}\right)^{-x_\phi},\\
&&\langle T(w)\rangle_a=0\,,\quad \langle T(z)\rangle_{{\rm ndl},a}={\textstyle{1\over 32}}\,\hat{c}\,z^{-2}\,,\\
&&\begin{array}{l}\langle T(w_1)T(w_2)\rangle_a^{{\rm cum}}={\textstyle{1\over 2}}\,\hat{c}\,(w_1-w_2)^{-4}\,,\\ 
\langle T(z_1)T(z_2)\rangle_{{\rm ndl},a}^{{\rm cum}}={\textstyle{1\over 32}}\,\hat{c}\,(z_1 z_2)^{-1}(z_1^{1/2}-z_2^{1/2})^{-4}\,,\end{array}
\label{avT1T2}
\end{eqnarray}
needed below.
Here the expressions $\langle ...\rangle_{{\rm ndl},a}$ for the  $z$ or needle geometry follow from the corresponding  $\langle ...\rangle_a$ in the upper half $w$ plane
and the conformal transformation properties of $\phi$ and $T$, shown  in and just above Eq.~(\ref{Ttransform}).

In analogy with Eq.~(\ref{kumu}) the two-point cumulant has the asymptotic form
\begin{equation}
\langle\phi(x,y)\Phi(X,Y)\rangle^{{\rm cum}}_{{\rm ndl,a}}\to \widehat{ F}_a^{(\phi)}(x,y)\Psi_a^{(\Phi)}(X,Y)\label{defsFPsi}
\end{equation}
for $z=x+iy=r^{i\vartheta}$ much closer to origin or tip of the needle than $Z=X+iY=Re^{i\Theta}$. The functions $\widehat{F}_a^{(\phi)}$ and $\Psi_a^{(\Phi)}$ can be determined as follows:
If $|z|\ll |Z|$, then $|w|\ll |W|$, and the boundary-operator expansion (\ref{BOE})  applies, except in the case $(\phi,a)=($order parameter, free$)$. For all other $(\phi,a)$ the corollary (\ref{avphiPhiabc}) of expansion (\ref{BOE}) leads to
\begin{eqnarray}
&&\langle\phi(u,v)\Phi(U,V)\rangle^{{\rm cum}}_a\to \mu_a^{(\phi)}v^{2-x_\phi}\langle T(u)\Phi(U,V)\rangle_a\nonumber\\
&&\to\left\{\begin{array}{l}{1\over 2}\,\hat c\, W^{-4}\,,\\
-2x_\Phi V^2 |W|^{-4}\langle \Phi(U,V)\rangle_a\,,\\ 
-2x_\phi v^2 W^{-4}\langle\phi(u,v)\rangle_a\,,\\
8\hat{c}^{-1} x_\phi x_\Phi v^2 V^2 |W|^{-4}\langle \phi(u,v)\rangle_a\langle \Phi(U,V)\rangle_a\,, \end{array}\right.
\begin{array}{l}
\phi=T,\; \Phi=T\\
\phi=T,\;\Phi\;{\rm primary}\\
\phi\;{\rm primary},\;\Phi=T\\
\phi\;{\rm and}\;\Phi\;{\rm primary}
\end{array}\,.\label{phiPhiwgeometry}
\end{eqnarray}
Here we have used Eqs.~(\ref{relationmuhphi}) and (\ref{avT1T2}) and evaluated $\langle T\phi\rangle_a$ and $\langle T\Phi\rangle_a$ using the Ward identity (\ref{GGWII}), as in footnote \cite{EEJune7}.
Conformally transforming Eq.~(\ref{phiPhiwgeometry}) to the needle geometry, we obtain
\begin{eqnarray}
&&\langle\phi(x,y)\Phi(X,Y)\rangle^{{\rm cum}}_{{\rm ndl},a}\nonumber\\
&&\to\left\{\begin{array}{l}
{1\over 32}\,\hat c\, (z Z^3)^{-1}\\
-{1\over 2}x_\Phi(zR)^{-1}\sin^2{\textstyle{\Theta\over 2}}\;\langle \Phi(X,Y)\rangle_{{\rm ndl},a}\\ 
-{1\over 2}x_\phi\,rZ^{-3}\sin^2{\textstyle{\vartheta\over 2}}\;\langle\phi(x,y)\rangle_{{\rm ndl},a}\\
8\hat{c}^{-1} x_\phi x_\Phi\, rR^{-1}\sin^2{\textstyle{\vartheta\over 2}}\;\sin^2{\textstyle{\Theta\over 2}}\;\langle \phi(x,y)\rangle_{{\rm ndl},a}\langle \Phi(U,V)\rangle_{{\rm ndl},a}
\end{array}\right.
\begin{array}{l}
\phi=T,\; \Phi=T\\
\phi=T,\;\Phi\;{\rm primary}\\
\phi\;{\rm primary},\;\Phi=T\\
\phi\;{\rm and}\;\Phi\;{\rm primary}
\end{array}\label{phiPhineedle}
\end{eqnarray}
Comparing Eq.~(\ref{defsFPsi}) and the first of Eqs.~(\ref{phiPhineedle}), we see that ${\widehat{F}^{(T)}}\propto z^{-1}$.  Choosing the arbitrary proportionality constant equal to 1, we find that
\begin{equation}
\begin{array}{l}
\widehat{F}_a^{(T)}(x,y)=z^{-1}\,,\\
\widehat{F}_a^{(\phi)}(x,y)=-16\hat{c}^{-1}x_\phi\, r\sin^2{\textstyle{\vartheta\over 2}}\,\langle\phi(x,y)\rangle_{{\rm ndl},a}\,,\quad\;\;\;\phi\;{\rm primary}\\
\Psi_a^{(T)}(X,Y)={\textstyle{1\over 32}}\,\hat c\,Z^{-3}\,,\\
\Psi_a^{(\Phi)}(X,Y)=-{\textstyle{1\over 2}\, }x_\Phi\, R^{-1}\sin^2{\textstyle{\Theta\over 2}}\;\langle \Phi(X,Y)\rangle_{{\rm ndl},a}\,.\;\;\;\Phi\;{\rm primary}
\end{array}\label{FandPsi}
\end{equation}
Equations~(\ref{defsFPsi}) and Eq.~(\ref{FandPsi}) determine  the asymptotic behavior of the two-point average $\langle\phi\Phi\rangle_{{\rm ndl},a}^{{\rm cum}}$ for $z$ much closer to the needle tip than $Z$. Here, as noted above, we exclude the case $(\phi,a)=$(order parameter,free).

As a check on these results, we have determined $\langle\phi\Phi\rangle_{{\rm ndl},a}^{\rm cum }$ exactly for the Ising model, conformally transforming the half-space results in Eqs.~(4.1) and (4.2) of Ref.~\cite{TWBX}. With $\phi$ and $\Phi=\sigma$ or $\epsilon$ and $a=+$ or $f$, there are 8 possibilities for $(\phi,\Phi,a)$. Two of these $(\sigma,\epsilon,f)$ and $(\epsilon,\sigma,f)$ are trivial, since the two-point average vanishes by symmetry, and for $(\sigma,\sigma,f)$ the prediction (\ref{defsFPsi}), (\ref{FandPsi}) of the boundary-operator expansion does not apply. In the other 5 cases, Eqs.~(\ref{defsFPsi}), (\ref{FandPsi}) and the asymptotic behavior of the exact Ising results are in complete agreement.

Finally, we note that the approach is easily adapted to the case of a needle with boundary condition $a$ on one side and $b$ on the other. On using the expansion Eq.~(\ref{kumu}) instead of (\ref{avphiPhiabc}), the first line of Eq.~(\ref{phiPhiwgeometry}) is replaced by
\begin{equation}
\langle\phi(u,v)\Phi(U,V)\rangle^{{\rm cum}}_{ab}\to F_{ab}^{(\phi)}\,(u-\zeta_1,v)\partial_{\zeta_1}\langle\Phi(U,V)\rangle_{ab}\,,
\end{equation}
and the steps that follow are modified accordingly.

\clearpage
\begin{figure}{Fig1}
\includegraphics[width=12cm]{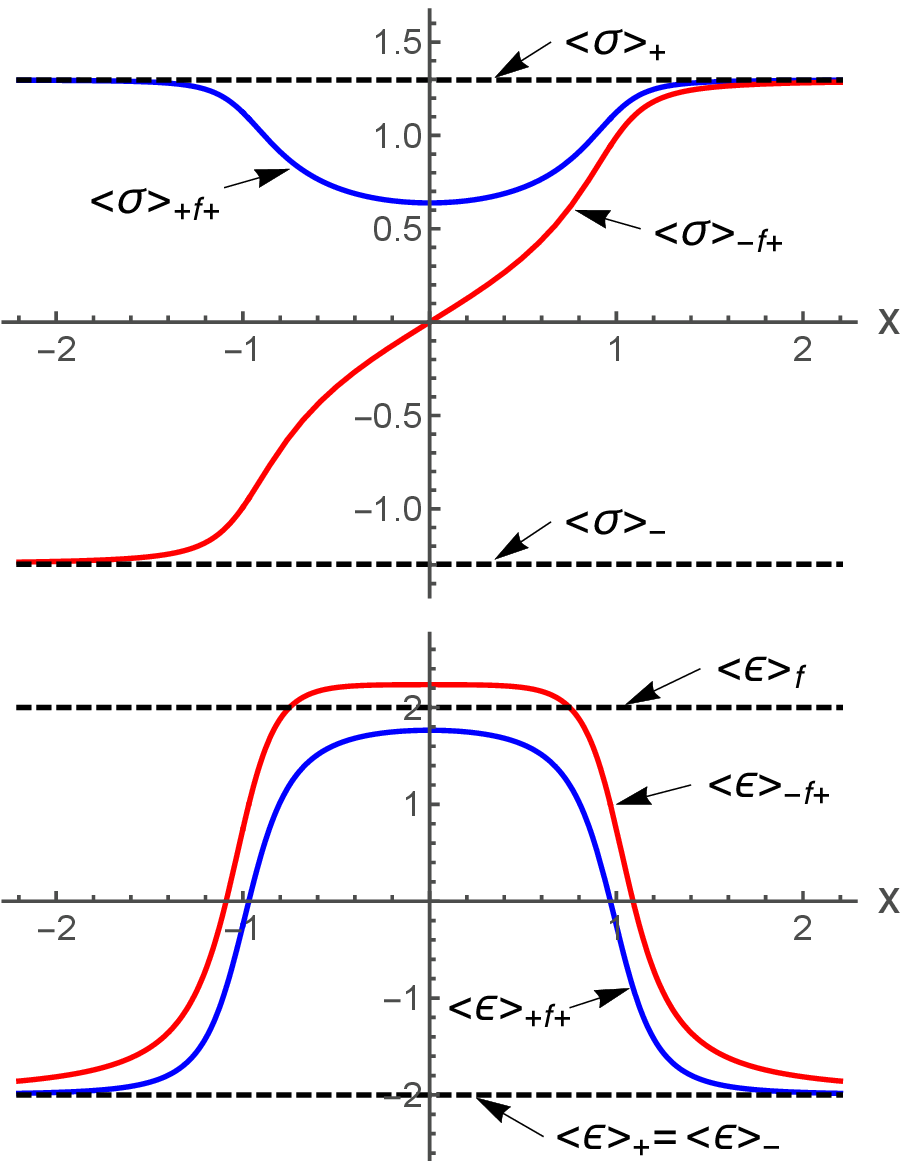}
\caption{Plots of $\langle\sigma\rangle$ and $\langle\epsilon\rangle$ for $+f+$ and $-f+$ boundaries, given in Eqs.~(\ref{sigpfp}),  (\ref{epspfp}), (\ref{sigmfp}), and (\ref{epsmfp}), as functions of $x$ for $y={1\over 4}$, $\zeta_1=-1$, and $\zeta_2=1$. The horizontal dashed lines indicate results for the uniform boundary conditions $+$, $-$, and $f$, given in Eqs.~(\ref{sigfixed}) and (\ref{epsfixedfree}). Since the $-f+$ boundary condition is less conducive to ordering than the  $+f+$ boundary condition, the curve for $\langle\epsilon\rangle_{-f+}$ lies above the curve $\langle\epsilon\rangle_{+f+}$ and, for $-{3\over 4}<x<{3\over 4}\,$, even above the dashed line representing $\langle\epsilon\rangle_f$.}
\label{fig1} 
\end{figure}


\newpage

\begin{thebibliography}{99}
\bibitem{BPZ} A. A. Belavin, A. M. Polyakov, and A. B. Zamolodchikov, Nuclear Phys. B {\bf 241}, 333 (1984).
\bibitem{CardyD-L} J. L. Cardy, in {\it Phase Transitions and Critical Phenomena}, edited by C. Domb and J. L. Lebowitz (Academic, New
York, 1986), Vol. 11, p. 55.
\bibitem{Cardyscp} J. L. Cardy, Nucl. Phys. B {\bf 240}, 514 (1984).
\bibitem{Cardytab} J. L. Cardy, Nucl. Phys. B {\bf 275}, 200 (1986); {\bf 324}, 581 (1989). 
\bibitem{TWBX} T. W. Burkhardt and T. Xue, Phys. Rev. Lett. {\bf 66}, 895 (1991); Nucl. Phys.  {\bf B354},653 (1991).
\bibitem{TWBG1} T. W. Burkhardt and I. Guim, Phys. Rev. B {\bf 36}, 2080 (1987).
\bibitem{TWBG2} T. W. Burkhardt and I. Guim, Phys. Rev. B {\bf 47}, 14 306 (1993).
\bibitem{EETWB} E. Eisenriegler and T. W. Burkhardt, Phys. Rev. E {\bf 94}, 032130 (2016).
\bibitem{SMED} A. Squarcini, A. Macio\l{}ek, E. Eisenriegler, and S. Dietrich, J. Stat. Mech. {\bf 2020}, 043208 (2020).
\bibitem{TWBEE} T. W. Burkhardt and E. Eisenriegler, Phys. Rev. Lett {\bf 74}, 3189 (1995); {\bf 78}, 2867 (1997); E. Eisenriegler and U. Ritschel,  Phys. Rev. B {\bf 51}, 13717 (1995).
\bibitem{BEK} G.Bimonte, T. Emig, and M. Kardar, Europhys. Lett. {\bf 104}, 21001 (2013). 
\bibitem{Diehl} H. W. Diehl, in {\em Phase Transitions and Critical Phenomena}, edited by C. Domb and J.L. Lebowitz (Academic, London, 1986), Vol.~{\bf 10}, p.~76;  H. W. Diehl, Int. J. Mod. Phys. B~{\bf 11}, 3503 (1997).
\bibitem{CardyLewellen} J. L. Cardy and D. C. Lewellen, Phys. Lett. B {\bf 259}, 274 (1991).
\bibitem{EEStap} E. Eisenriegler and M. Stapper, Phys. Rev. B 50, 10009 (1994).
\bibitem{Cardydistantwall} J. L. Cardy, Phys. Rev. Lett. {\bf 65}, 1443 (1990).
\bibitem{EE} E. Eisenriegler, J. Chem. Phys. {\bf 121}, 3299 (2004).
\bibitem{KC}  L. P. Kadanoff and H. Ceva, Phys. Rev. B {\bf 3}, 3918 (1971).
\bibitem{Cardydisorderop} J. L. Cardy, J. Phys. A  {\bf 17}, L961 (1984).
\bibitem{level2}
Differential equations (\ref{conformdiffeqsig})  and (\ref{diffeqsigmfp1}) follow from the conformal Ward identity for mixed boundary conditions $abc$... and the degeneracy of $\sigma$ at level two \cite{TWBX}. The average of the degeneracy condition is given by 
$\big\langle\left(-{4\over 3}\,\partial_z^2+L_{-2}\right)\sigma(z,\bar{z})\big\rangle_{abc\dots}^{(\zeta_1,\zeta_2,\dots)}= 0$, where $L_{-2}\,\sigma(z, \bar{z})=\int_C dz'(z'-z)^{-1}T(z')\sigma(z,\bar{z})$, and the integration path $C$ in the complex $z'$ plane encircles $z$ counterclockwise \cite{BPZ,CardyD-L}. Evaluating this average with $\langle T\sigma\rangle$ given by the Ward identity (\ref{GGWI}), except that $(z,Z,\Phi)$ are replaced by $(z',z,\sigma)$, leads to Eqs.~(\ref{conformdiffeqsig})  and (\ref{diffeqsigmfp1}). Equation (\ref{diffeqepsmfp1}) is obtained in a similar way.

\bibitem{G+R}I. S. Gradshteyn and I. M. Ryzhik {\it Table of Integrals, Series, and  Products}, AP New York and London 1965.
\bibitem{EEKD} E. Eisenriegler, M. Krech, and S. Dietrich, Phys. Rev. Lett. {\bf 70}, 619 (1993); {\bf 70}, 2051 (1993); Phys. Rev. B~{\bf 53}, 14377 (1996).
\bibitem{symarg} Multiplying Eq.~(\ref{BOE}) with $\phi(x_1,y_1)=\sigma_1$ by $\epsilon_2$, and averaging yields $\langle\sigma_1\epsilon_2\rangle_h-\langle\sigma_1\rangle_h\langle\epsilon_2\rangle_h\to\mu_h^{(\sigma)}y_1^{2-x_\sigma}\langle T(x_1)\epsilon_2\rangle_h$ for a system with a uniform $h$ boundary. Though correct for $h=+$ or $-$, this relation clearly does not apply for $h=f$, since the left-hand side vanishes everywhere in the half-plane, while the right side is non-vanishing.
\bibitem{McAvOs} D. M. McAvity and H. Osborn, Nucl. Phys. B {\bf 406}, 655 (1993).
\bibitem{DDE} H. W. Diehl, S. Dietrich, and E. Eisenriegler, Phys. Rev. B {\bf 27}, 2937 (1983).
\bibitem{TWBHWD} T. W. Burkhardt and H.W. Diehl, Phys. Rev. B {\bf 50}, 3894 (1994).
\bibitem{EEJune7} According to the boundary-operator expansion (\ref{BOE}), 
$$\langle T(x')\phi(z,\bar z)\rangle_a\to \mu_a^{(\phi)}y^{2-x_\phi}\langle T(x')T(x)\rangle_a=\mu_a^{(\phi)}y^{2-x_\phi}\,(\hat{c}/2)(x'-x)^{-4}$$
for a uniform boundary $a$ and $y\ll |x'-x|$. Substituting $\langle\phi\rangle_a\propto  (z-\bar z)^{-x_\phi}$ in the conformal Ward identity relating $\langle T\phi\rangle_a$ and $\langle\phi\rangle_a$ leads to a different expression
$$\langle T(x')\phi(z,\bar z)\rangle_a=-2x_\phi y^2 |x'-z|^{-4}\langle\phi\rangle_a\to -2x_\phi y^2(x'-x)^{-4}\langle\phi\rangle_a$$
for $y\ll |x-x'|$. Equating the rightmost terms in these two equations, we obtain Eq.(\ref{relationmuhphi}).
\bibitem{dilatation} For an $ab$ boundary, translational invariance parallel to the $x$ axis implies $\langle\phi(x,y)\rangle_{ab}^{(\zeta_1)}=\langle\phi(\delta x,y)\rangle_{ab}^{(0)}$, where $\delta x\equiv x-\zeta_1$. This and the invariance of $\lambda^{x_\phi}\langle\phi(\lambda\delta x,\lambda y)\rangle_{ab}^{(0)}$ under changes in the scale factor $\lambda$ lead to
$$\left(x_\phi+\delta x\partial_x+y\partial_y\right)\langle\phi(x,y)\rangle_{ab}^{(\zeta_1)}=
\left(x_\phi+\delta z\,\partial_z+\delta\bar z\,\partial_{\bar z}\right)\langle\phi(x,y)\rangle_{ab}^{(\zeta_1)}=0\,.$$ 
For an $abc$ boundary, analogous steps lead to 
$$\left(x_\phi+\delta z\,\partial_z+\delta\bar z\,\partial_{\bar z}+\delta\zeta\,\partial_{\zeta_2}\right)\langle\phi(x,y)\rangle_{abc}^{(\zeta_1,\zeta_2)}=0\,,\;\delta\zeta\equiv\zeta_2-\zeta_1\,.$$
\bibitem{FdG} M. E. Fisher and P.-G. de Gennes, C. R. Acad. Sci. Paris B {\bf 287}, 207 (1978).
\bibitem{RudnickJasnow} J. Rudnick and D. Jasnow, Phys. Rev. Lett. {\bf 49}, 1595 (1982).
\bibitem{Upton} Z. Borjan and P. J. Upton, Phys. Rev. Lett. {\bf 81}, 4911 (1998).
\end{thebibliography}
\end{document}